\documentclass[journal=jacsat,manuscript=article]{achemso}
\SectionNumbersOn
\usepackage{amsmath}
\usepackage[version=3]{mhchem} 
\usepackage[font=small,labelfont=bf]{caption}
\usepackage[font=small,labelfont=bf]{subcaption}
\usepackage{threeparttable}
\usepackage{booktabs,siunitx}
\usepackage{physics}
\usepackage{bm}
\usepackage{hyperref}

\author{Junhan Chen}
\affiliation[Upenn]
{Department of Chemistry, University of Pennsylvania, Philadelphia, Pennsylvania 19104, USA}
\author{Wenjie Dou}
\affiliation[1]
{School of Science, Westlake University, Hangzhou, Zhejiang 310024, China}
\alsoaffiliation[2]
{Institute of Natural Sciences, Westlake Institute for Advanced Study, Hangzhou, Zhejiang 310024, China}
\author{Joseph Subotnik}
\email{subotnik@sas.upenn.edu}
\phone{+1 (215) 746-7078}
\affiliation[Upenn]
{Department of Chemistry, University of Pennsylvania, Philadelphia, Pennsylvania 19104, USA}

\title[An \textsf{achemso} demo]
  {Active Spaces and Non-Orthogonal Configuration Interaction Approaches for Investigating Molecules on Metal Surfaces }

\abbreviations{IR,NMR,UV}
\keywords{American Chemical Society, \LaTeX}

\begin{document}

\begin{tocentry}

Some journals require a graphical entry for the Table of Contents.
This should be laid out ``print ready'' so that the sizing of the
text is correct.

Inside the \texttt{tocentry} environment, the font used is Helvetica
8\,pt, as required by \emph{Journal of the American Chemical
Society}.

The surrounding frame is 9\,cm by 3.5\,cm, which is the maximum
permitted for  \emph{Journal of the American Chemical Society}
graphical table of content entries. The box will not resize if the
content is too big: instead it will overflow the edge of the box.

This box and the associated title will always be printed on a
separate page at the end of the document.

\end{tocentry}

\begin{abstract}
    We test a set of multiconfigurational wavefunction approaches for calculating the ground state electron population for a two-site Anderson model representing a molecule on a metal surface. In particular, we compare ($i$) a Hartree Fock like wavefunction where frontier orbitals are allowed to be nonorthogonal versus $(ii)$ a fully non-orthogonal configuration interaction wavefunction based on  constrained Hartree-Fock states. We test both the strong and weak metal-molecule hybridization ($\Gamma$) limits as well as the strong and weak electron-electron repulsion ($U$) limits.
    We obtain accurate results  as compared with exact numerical renormalization group (NRG) theory, recovering charge transfer states where appropriate. The current framework should open a path  to run molecular non-adiabatic dynamics  on metal surfaces.  
\end{abstract}

\section{Introduction}

Solving embedding problem has attracted a great deal of attention in recent years \cite{kaneko2020charge, cevolani2018universal,han2018lattice,keshavarz2018electronic,knizia2012density,lee2019projection,bulik2014density,bulik2014electron,kluner2002periodic,sharifzadeh2008embedded,libisch2014embedded} as chemists look to exploring interesting interfacial phenomena, e.g., molecular resonance effects on metal surfaces \cite{gavnholt2008delta}, electron-coupled adsorption \cite{bunermann2015electron} and electron-coupled vibration \cite{morin1992vibrational, huang2000vibrational}. Embedding theory provides an attractive strategy to describe the electronic structure of extended systems (including interfacial systems) which can be impractical for traditional high accuracy quantum mechanics methods, e.g.
 coupled-cluster singles, doubles and full triples (CCSDT) \cite{raghavachari1989fifth}, second-order perturbation with complete active space (CASPT2) \cite{andersson1990p,andersson1992second} and  multireference configuration interaction (MRCI) \cite{bruna1987excited} or full configuration interaction (FCI) \cite{sherrill1999configuration,sherrill1996computational}. 

Historically, some of the earliest embedding calculations have addressed the Anderson Impurity model \cite{anderson1961localized} (or the more general Hubbard model).  By now, this model has been analzyed by a variety of exact impurity solvers including the numerical renormalization group (NRG) \cite{bulla2008numerical}, exact diagonalization (ED) \cite{fu2016numerical} and quantum monte carlo (QMC) \cite{gull2011continuous}. These benchmark  studies have then been very useful as far as benchmarking other, not exact but powerful, embedding methods, including  dynamical mean-field theory (DMFT) \cite{georges1996dynamical} and density matrix embedding theory (DMET) \cite{knizia2012density}.  To date, however, many of these powerful methods still have not been applied to study the problem of embedding realistic molecules on a realistic metal surface  where there are many two-electron matrix elements.  For such a mundane task,  constrained DFT  (CDFT) still remains the most practical approach, and the method  has been applied successfully to some extent \cite{kaduk2012constrained,ma2020pycdft,souza2013constrained,behler2007nonadiabatic}. That being said,  CDFT results can also be unreliable in some cases, e.g. strong molecule-metal coupling (i.e. strong hybridization) \cite{gavnholt2008delta} and fractional charge transfer \cite{mavros2015communication}.  Nonadiabatic dynamics remains just out of reach for many realistic potentials.

With this background in mind, the goal of this paper is to introduce an new electronic structure method designed specifically to address a molecule on a metal surface. Importantly, when the electron-electron repulsion energy on the impurity becomes large compared with the impurity-metal coupling (i.e. in the weak metal-molecule coupling limit), the true wavefunction will exhibit strong multireference character. In such a case, we must be able to capture the open-shell singlet character for an electron on or off the impurity; a single determinant wavefunction will not be a good reference in such a case. At the same time, of course, the correct wavefunction {\em will} exhibit simple single-reference character in the limit of strong metal-molecule coupling, and  a good solver must reduce to the RHF solution in such a case.  With these two limits in mind, below we explore a very basic approach whereby the frontier orbitals are modulated so that the total wavefunction will indeed contain open-shell character in the weak coupling limit and closed-shell character in the strong coupling limit. We call the resulting ansatz a closed-or-open shell Hartree Fock (COOS-HF) wavefunction.

In what follows, our specific goal will be to derive the necessary equations for solving for the COOS-HF equations just described, to apply this method to the Anderson Impurity model, and to compare our results to a constrained Hartree-Fock/Nonorthogonal configuration interaction (CHF/NOCI) formalism. We will show that surprisingly accurate results can be found with only a few variables to be  optimized.
We note that these results follow our previous work in embedding using configuration interaction spaces \cite{jin2020configuration,chen2021electronic}, but we emphasize that the present approach does not require diagonalizing a massive matrix of any kind. 

An outline of this manuscript is as follows. In Sec. \ref{sec:theory}, we introduce the COOS-HF formalism and we review the basic elements of a standard CHF/NOCI approach; we also include a brief comparison between COOS-HF and CASSCF(2,2) in subsection \ref{sec:theory}\ref{subsec:cooshfcas22}. In Sec. \ref{sec:results}, we show results for the ground state population  and energy, from which one can evaluate the accuracy of our approach. In Sec. \ref{sec:discussion}, we discuss spin-contamination and show that the present approach is free of any contamination. Furthermore, we contemplate different variations of the COOS-HF anasatz that one can imagine implementing (e.g.,  [\underline{p}artially \underline{o}ptimized] poCOOS-HF vs [\underline{f}ully \underline{o}ptimized] foCOOS-HF) and we place our results in the context of a CASSCF(2,2) formalism. We conclude in Sec. \ref{sec:conclusion}.

\section{Theory}
\label{sec:theory}
\subsection{The Model: Two-Site Anderson Impurity Model}
For this paper, our model of choice will be the two-site Anderson impurity model (AIM). 
Within a second quantized representation, the Hamiltonian  can be written as:
 \begin{equation}
 \begin{aligned}
 \label{eqn:model}
     \hat{H}&=\hat{H}_{one}+\hat{\Pi}\\
     \hat{H}_{one}&=\epsilon_{d_1}\sum_\sigma d_{1\sigma}^{\dagger}d_{1\sigma}+\epsilon_{d_2}\sum_\sigma d_{2\sigma}^{\dagger}d_{2\sigma}+t_d\sum_\sigma(d_{1\sigma}^{\dagger}d_{2\sigma}+d_{2\sigma}^{\dagger}d_{1\sigma})\\
     &+ \sum_{k\sigma}\epsilon_{k\sigma}c_{k\sigma}^{\dagger}c_{k\sigma}+\sum_{k\sigma}V_k(d_{1\sigma}^{\dagger}c_{k\sigma}+c_{k\sigma}^{\dagger}d_{1\sigma})\\
     \hat{\Pi}&=U(d_{1\uparrow}^{\dagger}d_{1\uparrow}d_{1\downarrow}^{\dagger}d_{1\downarrow}+d_{2\uparrow}^{\dagger}d_{2\uparrow}d_{2\downarrow}^{\dagger}d_{2\downarrow})
     \end{aligned}
 \end{equation}
 and the mean-field Fock operator (assuming spin restricted case) can be easily written as:
 \begin{equation}
 \label{eqn:F}
     \hat{F}=\hat{H}_{one}+U\expval{d_1^{\dagger}d_1}d_1^{\dagger}d_1+U\expval{d_2^{\dagger}d_2}d_2^{\dagger}d_2
 \end{equation}
 Here, $\{\hat{d_1}^\dagger,\hat{d_2}^\dagger\}$ refer to impurity atomic orbitals, the operators $\hat{c_k}^\dagger$ refer to bath (metal surface) atomic orbitals, and $\sigma$ 
refers to an electron spin.
$\epsilon_{d_1}, \epsilon_{d_2}$ and $\epsilon_k$ are one-electron ionization energies for the impurities and bath. $t_d$ is the hopping parameter between site 1 and site 2, $U$ represents the on-site coulomb repulsion for the impurity. $V_k$ represents the hybridization between impurity site 1 and the metal bath, and as in the wide band approximation, is characterized by:
\begin{equation}
    \Gamma=\Gamma(\epsilon)=2\pi\sum_k |V_k|^2\delta(\epsilon-\epsilon_k) ,
\end{equation}
where $\Gamma$ is assumed to be constant through the whole energy spectrum $\epsilon$.

The Hamiltonian in Eq. \ref{eqn:model} characterizes many different physical processes because, in certain parameter regimes, one can identify states with effectively open shell singlet character within the set of impurity orbitals;  in other regimes, one can identify charge transfer between impurity and metal; and of course  these two phenomena cannot be fully disentangled for all parameter regimes, highlighting a more realistic version of electron-electron correlation than is found for a single-site Anderson Holstein Hamiltonian.

We will now address the two wavefunction approaches that we wish to compare as far as assessing the ground-state electronic structure for the system plus bath: a constratined Hartree Fock configuration interaction  approach versus a closed-or-open-shell frontier orbital wavefunction approach.

\subsection{Method 1: Constrained Hartree-Fock Based Nonorthogonal Configuration Interaction (CHF/NOCI)  }

\subsubsection{Constrained HF states}
Following  Ref. \cite{wu2005direct}, a general constraint to the density can often be written as,
\begin{equation}
    \sum_{\sigma}\int w_c^{\sigma}(\bm{r})\rho^{\sigma}(\bm{r})d\bm{r}=N_c
\end{equation}
where $w_c(\bm{r})$ acts as a weight function that defines the constrained property, $\sigma$ represents the electron spin, $\rho(\bm{r})$ represents the charge density and $N_c$ represents the total number of constrained charge (or electrons). By adding one Lagrange multiplier, $V_c$, one can optimize a general functional of the density $E\left[ \rho \right]$ with a prescribed constraint by looking for extrema of the following function:
\begin{equation}
\label{eqn:W}
    W[\rho,V_c]=E[\rho]+V_c(\sum_{\sigma}\int w_c^{\sigma}(\bm{r})\rho^{\sigma}(\bm{r})d\bm{r}-N_c)
\end{equation}
Within constrained density functional theory\cite{wu2005direct}, one looks to minimize the energy functional with the constraint that orbitals are normalized, and one then arrives as the standard constrained DFT equations:
\begin{equation}
    \qty(-\frac{1}{2}\laplacian+v_n(\bm{r})+\int \frac{\rho(\bm{r'})}{|\bm{r}-\bm{r'}|}d\bm{r'}+v_{xc\sigma}(\bm{r})+V_cw_c^{\sigma}(\bm{r}))\psi_{i\sigma}=\epsilon_{i\sigma}\psi_{i\sigma}
\end{equation}

This language (based on continuous real space or plane wave basis sets) can easily be extended to the realm of discretized site basis sets as appropriate for a generalized Anderson model  under a restricted Hartree-Fock (RHF) framework.
The relevant density constraint (defining the number of electrons on the impurity) becomes
\begin{equation}
    \Tr(\hat{\rho}\hat{w}_c)=N_c
\end{equation}
where the density matrix $\hat{\rho}$ and the weight matrix $\hat{w_c}$ are defined as:
\begin{equation}
\label{eqn:rho}
    \hat{\rho}=\sum_{i\in occ}\ketbra{\psi_i}{\psi_i}
\end{equation}
\begin{equation}
    \hat{w}_c=\sum_{\mu\in impurity}w_{\mu}\ketbra{d_\mu}{d_\mu}
\end{equation}
Here, the set $\left\{\ket{\psi_i}\right\}$ are the eigenvectors of the Hartree-Fock equation and we set all site weights as $w_\mu = 1$. Specifically speaking, the weight matrix for two-site Anderson model will be:
\begin{equation}
\label{eqn:w}
    \hat{w}_c=\ketbra{d_1}{d_1}+\ketbra{d_2}{d_2}
\end{equation}
We will explore all possible integer values for the  constrained number of integer electrons on the two impurity sites:
\begin{equation}
\label{eqn:Nc}
    N_c\in \{0e,1e,2e,3e,4e\}
\end{equation}

The final constrained Hartree-Fock equation can be written as
\begin{equation}
\label{eqn:CHF}
    \qty( \hat{F} + V_c\hat{w}_c)\psi_i=\epsilon_i\psi_i,
\end{equation}
where the fock matrix $\hat{F}$ is defined in Eq. \ref{eqn:F}. Note that even though $N_c$ is not explicitly included in Eq. \ref{eqn:CHF},  a value of $N_c$ is required to find the optimized Lagrange multiplier, $V_c$.

Here we use the first derivative and second derivative\cite{wu2005direct} of $W(V_c)$ (assuming restricted orbitals) in Eq. \ref{eqn:W}: 
\begin{align}
\label{eqn:dW}
    \dv{W}{V_c}
    =\sum_{\sigma}\int{w_c^{\sigma}(\bm{r})\rho^{\sigma}(\bm{r})}d\bm{r}-N_c=2\mel{d_1}{\hat{\rho}}{d_1}+2\mel{d_2}{\hat{\rho}}{d_2}-N_c\\
    \label{eqn:d2W}
    \dv[2]{W}{V_c}=2\sum_{\sigma}\sum_i^{N_{\sigma}}\sum_{a>N_{\sigma}}\frac{|\mel{\psi_{i\sigma}}{w_c^{\sigma}}{\psi_{a\sigma}}|^2}{\epsilon_{i\sigma}-\epsilon_{a\sigma}}=4\sum_{i\in{occ}}\sum_{a\in{vir}}\frac{|\mel{\psi_{i}}{w_c}{\psi_{a}}|^2}{\epsilon_{i}-\epsilon_{a}}
\end{align}
Here $\hat{\rho}$ is the density matrix operator defined in Eq. \ref{eqn:rho}, $d_1$ and $d_2$ represent two impurity sites, $w_c$ is the weight matrix defined in Eq. \ref{eqn:w}, $\{\psi_{i}\}$ or $\{\psi_{a}\}$ and $\epsilon_i$ or $\epsilon_a$ are eigenvectors and eigenvalues of Eq. \ref{eqn:CHF}, respectively. As discussed in Ref.\cite{wu2005direct} , Eq. \ref{eqn:d2W} is derived from simple first-order perturbation theory of the Kohn-Sham equations. An algorithm to determine the constrained Hartree-Fock solution can be summarised as follows:
\begin{enumerate}
    \item Choose the desired value of $N_c$ in Eq. \ref{eqn:Nc};
    \item Guess a Lagrange multiplier, e.g. $V_c=0$;
    \item Guess an impurity population $\expval{d_1^{\dagger}d_1}=\expval{d_2^{\dagger}d_2}=\frac{N_c}{4}$;
    \item Construct the Fock matrix using Eq. \ref{eqn:F};
    \item Solve Eq. \ref{eqn:CHF} self-consistently and obtain eigenvectors $\{\psi\}$ and eigenvalues $\{\epsilon\}$;
    \item Calculate a new impurity population $\expval{d_1^{\dagger}d_1}$ and $\expval{d_2^{\dagger}d_2}$;
    \item Take a Newton step to calculate a new $V_c$ by Eqs. \ref{eqn:dW}-\ref{eqn:d2W};
    \item Repeat Step 4-7 until $V_c$ converges.
\end{enumerate}

\subsubsection{Non-Orthogonal Configuration Interaction}
The total number of electrons on the two impurity sites is required to be an integer between 0 and 4. Therefore, one naturally finds five diabatic constrained Hartree-Fock (CHF) states for our model problem: $\ket{\Phi_{4e}}, \ket{\Phi_{3e}},\ket{\Phi_{2e}},\ket{\Phi_{1e}},\ket{\Phi_{0e}}$, respectively. Since these configurations are not orthogonal to each other, a configuration interaction Hamiltonian can be constructed as (see Appendix \ref{sec:appendix}\ref{sec:nomel} for a detailed derivation of the necessary matrix elements),
\begin{equation}
    H^{CHF/NOCI}=
    \begin{bmatrix}
    \mel{\Phi_{4e}}{H}{\Phi_{4e}} & \mel{\Phi_{4e}}{H}{\Phi_{3e}} & \mel{\Phi_{4e}}{H}{\Phi_{2e}} & \mel{\Phi_{4e}}{H}{\Phi_{1e}} & \mel{\Phi_{4e}}{H}{\Phi_{0e}}\\
    \mel{\Phi_{3e}}{H}{\Phi_{4e}} & \mel{\Phi_{3e}}{H}{\Phi_{3e}} & \mel{\Phi_{3e}}{H}{\Phi_{2e}} & \mel{\Phi_{3e}}{H}{\Phi_{1e}} & \mel{\Phi_{3e}}{H}{\Phi_{0e}}\\
    \mel{\Phi_{2e}}{H}{\Phi_{4e}} & \mel{\Phi_{2e}}{H}{\Phi_{3e}} & \mel{\Phi_{2e}}{H}{\Phi_{2e}} & \mel{\Phi_{2e}}{H}{\Phi_{1e}} & \mel{\Phi_{2e}}{H}{\Phi_{0e}}\\ 
    \mel{\Phi_{1e}}{H}{\Phi_{4e}} & \mel{\Phi_{1e}}{H}{\Phi_{3e}} & \mel{\Phi_{1e}}{H}{\Phi_{2e}} & \mel{\Phi_{1e}}{H}{\Phi_{1e}} & \mel{\Phi_{1e}}{H}{\Phi_{0e}}\\
    \mel{\Phi_{0e}}{H}{\Phi_{4e}} & \mel{\Phi_{0e}}{H}{\Phi_{3e}} & \mel{\Phi_{0e}}{H}{\Phi_{2e}} & \mel{\Phi_{0e}}{H}{\Phi_{1e}} & \mel{\Phi_{0e}}{H}{\Phi_{0e}}\\
    \end{bmatrix}
\end{equation}
and the corresponding overlap $S$ is expressed as:
\begin{equation}
    S=
    \begin{bmatrix}
    1 & \braket{\Phi_{4e}}{\Phi_{3e}}& \braket{\Phi_{4e}}{\Phi_{2e}}& \braket{\Phi_{4e}}{\Phi_{1e}}& \braket{\Phi_{4e}}{\Phi_{0e}} \\
    \braket{\Phi_{3e}}{\Phi_{4e}} & 1 & \braket{\Phi_{3e}}{\Phi_{2e}} & \braket{\Phi_{3e}}{\Phi_{1e}}& \braket{\Phi_{3e}}{\Phi_{0e}} \\
    \braket{\Phi_{2e}}{\Phi_{4e}} & \braket{\Phi_{2e}}{\Phi_{3e}} & 1 & \braket{\Phi_{2e}}{\Phi_{1e}} & \braket{\Phi_{2e}}{\Phi_{0e}} \\
    \braket{\Phi_{1e}}{\Phi_{4e}} & \braket{\Phi_{1e}}{\Phi_{3e}} & \braket{\Phi_{1e}}{\Phi_{2e}} & 1 & \braket{\Phi_{1e}}{\Phi_{0e}} \\
    \braket{\Phi_{0e}}{\Phi_{4e}} & \braket{\Phi_{0e}}{\Phi_{3e}} & \braket{\Phi_{0e}}{\Phi_{2e}} & \braket{\Phi_{0e}}{\Phi_{1e}} & 1 \\
    \end{bmatrix}
\end{equation}
To transform the interaction Hamiltonian into an orthogonal configuration basis, a transformation matrix $X$ will be introduced, which satisfies:
\begin{equation}
    X^{\dagger}SX=I
\end{equation}
In this paper, X is chosen as: $X=S^{-\frac{1}{2}}$. By diagonalizing the matrix $X^{\dagger}H^{CHF/NOCI}X$, one obtains five adiabatic constrained Hartree-Fock nonorthogonal configuration interaction (CHF/NOCI) states. 
\subsection{Method 2: Closed or Open Shell Hartree-Fock (COOS-HF) }
Having discussed a standard CHF/NOCI approach to charge transfer, let us now introduce an alternative  multiconfigurational approach that is not based on a configuration interaction Hamiltonian but rather on a set of frontier (non-orthogonal) orbitals. 
Our approach is to consider a wavefunction of the following form:
\begin{equation}
\label{eqn:cooshf}
    \ket{\Phi_{COOS-HF}}=\frac{1}{\sqrt{N}}(\ket{\Phi_{pq}}+\ket{\Phi_{qp}}),
    \end{equation}
where $N$ is the normalization factor: 
\begin{equation}
N=2(1+\braket{p}{q}^2),
\end{equation}
$\bm{p},\bm{q}$ are two nonorthogonal orbitals and $\ket{\Phi_{pq}},\ket{\Phi_{qp}}$ are the following two slater determinants for a $2n-$electrons system:
\begin{equation}
\label{eqn:pq}
\ket{\Phi_{pq}}=
    \begin{vmatrix}
    \psi_1(1) & \bar{\psi_1}(1) & \cdots  & p(1) & \bar{q}(1) \\
    \psi_1(2) & \bar{\psi_1}(2) & \cdots  & p(2) & \bar{q}(2) \\
    \vdots & \vdots & \ddots  & \vdots & \vdots \\
    \psi_1(2n) & \bar{\psi_1}(2n) & \cdots  & p(2n) & \bar{q}(2n) \\
    \end{vmatrix}
\end{equation}
\begin{equation}
\label{eqn:qp}
\ket{\Phi_{qp}}=
    \begin{vmatrix}
    \psi_1(1) & \bar{\psi_1}(1) & \cdots  & q(1) & \bar{p}(1) \\
    \psi_1(2) & \bar{\psi_1}(2) & \cdots  & q(2) & \bar{p}(2) \\
    \vdots & \vdots & \ddots  & \vdots & \vdots \\
    \psi_1(2n) & \bar{\psi_1}(2n) & \cdots  & q(2n) & \bar{p}(2n) \\
    \end{vmatrix}
\end{equation}
For the moment, we make no stipulations about the character of the orbitals $\bm{p}$ and $\bm{q}$; they could be parallel or they could be orthogonal.   And so, we refer to the wavefunction in Eq. \ref{eqn:cooshf} as a closed-or-open shell HF ansatz.

Two options are now possible as far optimizing such a COOS-HF wavefunction. First, one could imagine optimizing all of the orbitals, $\{\psi_1,\psi_2,...,\bm{p},\bm{q}\}$. We will call this approach a \underline{f}ully \underline{o}ptimized COOS-HF (foCOOS-HF) ansatz. 

The second approach is simpler and is based on a closed shell RHF reference state.
In such a case, we can seek three nonorthogonal orbitals $\bm{o},\bm{p},\bm{q}$, so that a simplified \underline{p}artialy \underline{o}ptimized COOS-HF (poCOOS-HF) wavefunction can be written as:
\begin{equation}
\label{eqn:pocooshf}
    \ket{\Phi_{poCOOS-HF}}=\frac{1}{\sqrt{N}}(\ket{\Psi_{o\bar{o}}^{p\bar{q}}}+\ket{\Psi_{o\bar{o}}^{q\bar{p}}}),
\end{equation}
In Eq. \ref{eqn:pocooshf}, the choice of orbitals $\bm{o}$, $\bm{p}$, and $\bm{q}$ is critical. For this wavefunction, we insist that orbital $\bm{o}$ must be an occupied orbital, but we make no such assumption about orbitals $\bm{p}$ and $\bm{q}$ (our ``active'' orbitals) -- even though p and q have been written for convenience in the superscript of the kets $\ket{\Phi_{o\bar{o}}^{p\bar{q}}}$ and $\ket{\Phi_{o\bar{o}}^{q\bar{p}}}$.   More precisely, orbital $\bm{p}$ need not be orthogonal to orbital $\bm{o}$ or orbital $\bm{q}$. Thus, when we optimize Eq. \ref{eqn:pocooshf} for orbitals $\bm{o}$, $\bm{p}$, and $\bm{q}$, we can indeed recover the starting RHF ansatz by  picking $\bm{o}$=$\bm{p}$=$\bm{q}$.

In particular, we can parameterize the spatial components of $\bm{o}$,$\bm{p}$, and $\bm{q}$ in the basis of RHF canonical occupied orbitals $\{o_i\}$ and virtual orbitals $\{v_a\}$:
\begin{equation}
\label{eqn:opq}
\begin{aligned}
    \ket{o}&=\sum_i c_i\ket{o_i} \\
    \ket{p}&=d_o\ket{o}+\sum_a d_a\ket{v_a}\\
    \ket{q}&=\tilde{d}_o\ket{o}+\sum_b \tilde{d}_b\ket{v_b}
\end{aligned}    
\end{equation}
Note that, if we define the core orbitals to be all of those occupied orbitals $\bm{i}$ orthogonal to orbital $\bm{o}$,   the following identities also hold:
\begin{align}
    \braket{i}{o}&=0\\
    \braket{i}{p}&=0\\
    \braket{i}{q}&=0
\end{align}
and core orbitals $\bm{i}$ together with orbital $\bm{o}$ forms the RHF occupied space, i.e.:
\begin{equation}
    \sum_i\ket{i}\bra{i}+\ket{o}\bra{o}=\sum_i\ket{o_i}\bra{o_i}
\end{equation}


\subsubsection{COOS-HF as a subset of a CASSCF(2,2) calculation}
\label{subsec:cooshfcas22}
The astute reader will notice that the COOS-HF wavefunction ansatz in Eq. \ref{eqn:cooshf} represents a singlet configuration without spin contamination and that the ansatz is clearly a subset of a CASSCF(2,2) ansatz by writing nonorthogonal orbitals $\bm{p},\bm{q}$ in the orthonormal basis $\bm{a},\bm{b}$.
\begin{equation}
\begin{aligned}
    \ket{p}&=\cos\theta \ket{a}+\sin\theta\ket{b}\\
    \ket{q}&=\cos\eta\ket{a}+\sin\eta\ket{b}
\end{aligned}
\end{equation}
Therefore, the wavefunction in Eq. \ref{eqn:cooshf} can be
 written as (for simplicity, core orbitals are ignored):
\begin{equation}
    \frac{\ket{p\bar{q}}+\ket{q\bar{p}}}{\sqrt{N}}=\frac{1}{\sqrt{N}}\{2\cos\theta\cos\eta\ket{a\bar{a}}+2\sin\theta\sin\eta\ket{b\bar{b}}+(\cos\theta\sin\eta+\cos\eta\sin\theta)(\ket{a\bar{b}}+\ket{b\bar{a}})\}
\end{equation}
This equation is clearly of the CASSCF(2,2) form (where the CI coefficients would normally be written as),
\begin{equation}
    \alpha\ket{a\bar{a}}+\beta\ket{b\bar{b}}+\gamma(\ket{a\bar{b}}+\ket{b\bar{a}})
\end{equation}
if one makes the substitution: 
\begin{equation}
\label{eqn:CI coeffs}
    \begin{aligned}
    \alpha&=\frac{2\cos\theta\cos\eta}{\sqrt{N}}\\
    \beta&=\frac{2\sin\theta\sin\eta}{\sqrt{N}}\\
    \gamma&=\frac{(\cos\theta\sin\eta+\cos\eta\sin\theta)}{\sqrt{N}}\\
    N&=2*[1+(\cos\theta\cos\eta+\sin\theta\sin\eta)^2]
    \end{aligned}
\end{equation}
 A more complete discussion of this correspondence will be given in the discussion section.
\subsubsection{Solving for the poCOOS-HF orbitals and energy}
Let us now discuss how one can most easily solve for the poCOOS-HF orbitals. For the ansatz in Eq. \ref{eqn:pocooshf}, the expectation value for the   total energy is: 
\begin{equation}
    \mel{\Phi_{poCOOS-HF}}{H}{\Phi_{poCOOS-HF}}=\frac{\mel{\Psi_{o\bar{o}}^{p\bar{q}}}{H}{\Psi_{o\bar{o}}^{p\bar{q}}}+\mel{\Psi_{o\bar{o}}^{p\bar{q}}}{H}{\Psi_{o\bar{o}}^{q\bar{p}}}}{1+\braket{p}{q}^2}
    \label{eqn:Ercs}
\end{equation}
where
\begin{equation}
\begin{aligned}
\mel{\Psi_{o\bar{o}}^{p\bar{q}}}{H}{\Psi_{o\bar{o}}^{p\bar{q}}}&=\mel{\cdots i \cdots p \cdots \bar{i} \cdots \bar{q}}{H}{\cdots i \cdots p \cdots \bar{i} \cdots \bar{q}}\\
&=2h_{ii}+h_{pp}+h_{qq}+(\sum_i ii + pp|\sum_j jj + qq)
\end{aligned}
\end{equation}
\begin{equation}
\begin{aligned}
\mel{\Psi_{o\bar{o}}^{p\bar{q}}}{H}{\Psi_{o\bar{o}}^{q\bar{p}}}&=\mel{\cdots i \cdots p \cdots \bar{i} \cdots \bar{q}}{H}{\cdots i \cdots q \cdots \bar{i} \cdots \bar{p}}\\
&=2h_{ii}\braket{p}{q}^2+2h_{pq}\braket{p}{q}+(\sum_i ii\braket{p}{q} + pq|\sum_j jj\braket{p}{q} + pq)
\end{aligned}
\end{equation}
Furthermore, this expression can be simplified using $E_{RHF}$ as a reference. Recall that the RHF energy is
\begin{align}
E_{RHF}=2h_{ii}+\sum_{ij}(ii|jj)+2h_{oo}+2\sum_{i}(ii|oo)+(oo|oo),
\end{align}
If we then define Fock operators as:
\begin{equation}
\begin{aligned}
    f_{oo}=h_{oo}+\sum_i (ii|oo) + (oo|oo)\\
    f_{pp}=h_{pp}+\sum_i (ii|pp) + (oo|pp)\\
    f_{qq}=h_{qq}+\sum_i (ii|qq) + (oo|qq)\\
    f_{pq}=h_{pq}+\sum_i (ii|pq) + (oo|pq)
\end{aligned}
\end{equation}
then the poCOOS-HF energy becomes:
\begin{equation}
\begin{aligned}
\label{eqn:Epocoos}
\mel{\Phi_{poCOOS-HF}}{H}{\Phi_{poCOOS-HF}}&=E_{RHF}-2f_{oo}+(oo|oo)\\
&+\frac{1}{1+\braket{p}{q}^2}[f_{pp}+f_{qq}+2f_{pq}\braket{p}{q}+2(pp|qq)]\\
&-\frac{1}{1+\braket{p}{q}^2}[(oo|pp)+(oo|qq)+2(oo|pq)\braket{p}{q}]
\end{aligned}
\end{equation}
Once the gradient is obtained (see Appendix\ref{sec:appendix}\ref{sec:gradient} for a complete derivation of the analytical energy gradient), one can use a lagrange multiplier and a quasi-Newton method to minimize the objective function. To write the equations more succinctly, let us use the symbol $x$  to represent a generic variable in the poCOOS-HF variable space in Eq. \ref{eqn:opq} (all of whose variables at denoted capital $X$): 
\begin{equation}
X=\{\{c_i\},d_o,\{d_a\},\tilde{d_o},\{\tilde{d_b}\}\}
\end{equation}
The lagrangian operator can then be written as:
\begin{equation}
    \min_{x\in X} \mathcal{L}(x)=E_{poCOOS-HF}(x)-\lambda_1C_1(x)-\lambda_2C_2(x)-\lambda_3C_3(x),
\end{equation}
where $\lambda_1, \lambda_2, \lambda_3$ are lagrange multipliers and the three constraints are:
\begin{equation}
    \begin{aligned}
        C_1(x)&=\sum_ic_i^2-1\\
        C_2(x)&=d_o^2+\sum_ad_a^2-1\\
        C_3(x)&=\tilde{d}_o^2+\sum_b\tilde{d}_b^2-1
    \end{aligned}
\end{equation}
Just as one would solve for an unconstrained objective function, we use a Newton iteration to solve for the present lagrangian (the  so-called Newton-KKT equation) \cite{wright1999numerical}:
\begin{equation}
    \begin{bmatrix}
        \grad^2_{xx}\mathcal{L}_k & -\grad C_1(x_k) & -\grad C_2(x_k) & -\grad C_3(x_k)\\
        \grad C_1^T(x_k) &0 &0 &0 \\
        \grad C_2^T(x_k)&0 &0 &0 \\
        \grad C_3^T(x_k) &0 &0 &0 \\
    \end{bmatrix}
    \begin{bmatrix}
        p_k\\
        \lambda_{1k}\\
        \lambda_{2k}\\
        \lambda_{3k}
    \end{bmatrix}
    =
    \begin{bmatrix}
        -\grad E_{poCOOS-HF}(x_k)\\
        -C_1(x_k)\\
        -C_2(x_k)\\
        -C_3(x_k,)\\
    \end{bmatrix}
\end{equation}
Here, $p_k$ is the walking direction for $x_k$, i.e. $x_{k+1}=x_k+\alpha p_k$ and the step length $\alpha$ is obtained from a  line search. To reduce the computational cost, the hessian $\grad^2_{xx}\mathcal{L}$ is also approximated and updated by a BFGS scheme. In practice, for the problems below and with a reasonable starting guess, we require roughly ten cycles (i.e. line searches).

 \section{Results and discussion}
 \label{sec:results}
In this paper, our goal is to compare the ground state properties (electron population and energy) as predicted by the methods (CHF/NOCI and poCOOS-HF) above and to assess their power for propagating adiabatic dynamics; in a future publication, we will address excited state properties (and e.g., we will benchmark against the ROKS method\cite{filatov1999spin,kowalczyk2013excitation}) so that we can assess running nonadiabatic dynamics.
For the present case, because we focus on ground state theory, it is fairly straightforward to obtain exact benchmark energies using 
numerical renormalization group theory (NRG), which recovers only electron population (not total energy which would depend on the number of discrete orbitals in the bath). We will also benchmark our results against
RHF and UHF.

Because molecules are very diverse and their properties  can cover a multitude of chemisorption and physisorption regimes, we will test the algorithms above in three different onsite repulsion regimes: weak metal-molecule coupling $U=10\Gamma$, intermediate coupling $U=5\Gamma$ and strong coupling $U=\Gamma$. 
 Fig. \ref{fig:n_imp} plots the ground state spin-up electron population on  impurity site 1.
We set the two impurity energies equal to each other, i.e. $\epsilon_{d} \equiv \epsilon_{d_1} = \epsilon_{d_2}$,
This onsite energy $\epsilon_d$ (which is varied along the $x-$axis) can be considered a charge transfer coordinate.
Each plot is separated into two $\epsilon_d$ regimes, where the total number of electrons on the  impurities range from 4 electrons to 2 electrons and 2 electrons to 0 electron, respectively; note that 
$\expval{n_{tot}}=4\expval{n_{1\uparrow}}$ for this restricted system.
In the following context, these two regimes will be represented as $4\geq\expval{n_{tot}}\geq2$ and $2\geq\expval{n_{tot}}\geq0$. The black line is the exact Numerical Renormalization Group (NRG) results for benchmark. The light blue line is restricted Hartree-Fock (RHF). The dark blue line is the unrestricted Hartree-Fock (UHF) result. 
The red line is constrained Hartree-Fock with non-orthogonal configuration interaction (CHF/NOCI). And lastly, the green line is the partially optimized closed-or-open shell Hartree-Fock (poCOOS-HF) result.

To begin with, consider the performance of the RHF and UHF methods. As one can see in Fig. \ref{fig:n_imp}(c), for the strong coupling regime, where static correlation is minimal, RHF itself is already a reasonable approximation to the ground state. However, for the weak or intermediate metal-molecule coupling regime, RHF becomes qualitatively incorrect when static correlation begins to dominate.   At this point, UHF does agree pretty well with NRG -- but with two obvious disadvantages.  First, there is a large discontinuity in the UHF results at the Coulson-Fischer point (see Figs. \ref{fig:n_imp}(a,b)).  Second, there is  spin-contamination problem (which will be discussed in  detail in Sec.
\ref{sec:discussion}\ref{subsec:spincontamination}).

Next, we turn to the CHF/NOCI method. As is well known, one of the disadvantages of CHF(or CDFT) is that the method can fail to describe strongly coupled molecule-metal systems where the subsystem being constrained (in this paper, the two impurity sites ) and the unconstrained subsystem (in this paper, the bath) are difficult to distinguish. Thus, the most important test of such a system for CHF/CI will be the strong hybridization case.  As can be seen in Fig. \ref{fig:n_imp}, while CHF/NOCI (the red line) matches up with NRG pretty well in the weak coupling ($U=10\Gamma$) and intermediate coupling ($U=5\Gamma$) regime,  the method fails in the strong coupling ($U=\Gamma$) regime.

Lastly, let us address the green curve in Fig. \ref{fig:n_imp}, representing the partially optimized closed-or-open shell Hartree Fock (poCOOS-HF) results. As can be seen in Fig. \ref{fig:n_imp}, poCOOS-HF results match with NRG results fairly well in the three different coupling regimes. As one would hope, in the weak and intermediate coupling regimes, poCOOS-HF behaves like UHF but with no spin-contamination; whereas in strong coupling regime, poCOOS-HF follows the (smooth) RHF solution. As a side note, we mention that, in Fig. \ref{fig:n_imp}(c), where the coupling is so strong that RHF and UHF can be considered close to the exact solution, most of the small offset between the RHF/UHF/ poCOOS-HF results and thne NRG results can be attributed to the small systematic error of the NRG method  (related to the choice of chain length, logarithmic discretization parameter, temperature parameter and energy truncation \cite{bulla2008numerical}).  (For instance, the NRG approach will not be exactly on top of the RHF line even for $\Gamma = 0$, where RHF is truly exact.)
These results are encouraging for future dynamics simulations.

Next, we consider energies.
Fig. \ref{fig:energy} plots the ground state energy (relative to RHF) as calculated by UHF, CHF/NOCI and poCOOS-HF in three different coupling regimes: (a) $U=10\Gamma$, (b) $U=5\Gamma$ and (c) $U=\Gamma$. The x axis is the same as in  Fig. \ref{fig:n_imp}. One can see from Fig. \ref{fig:energy} that the poCOOS-HF energy is very close to UHF and even lower than UHF when $U=5\Gamma$. Both UHF and poCOOS-HF give a maximum energy correction when $\expval{n_{tot}}=3$ and $\expval{n_{tot}}=1$, in which case the impurity has open-shell singlet character. Interestingly, in Fig. \ref{fig:energy}(a), even though the CHF/NOCI energy is about 1e-3 higher than UHF or poCOOS-HF, the CHF/NOCI electron population (in Fig. \ref{fig:n_imp}(a)) is still pretty close to the UHF or poCOOS-HF results. 


\begin{figure}[htbp]
\centering

\begin{subfigure}[t]{\textwidth}
\centering
\vspace{-50pt}
\hspace*{-0mm}\includegraphics[width=1.0\linewidth]{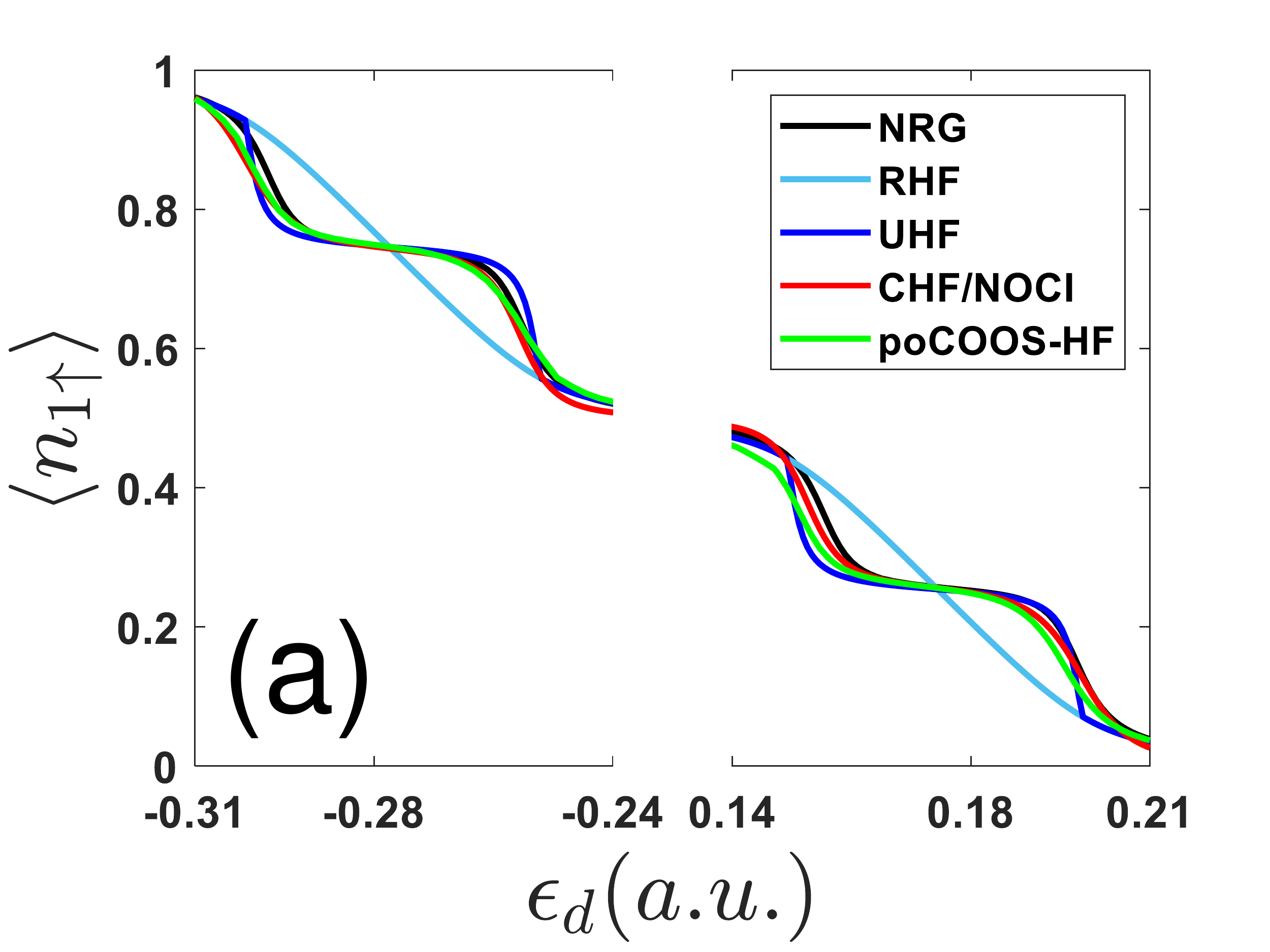}
\end{subfigure}
\begin{subfigure}[t]{\textwidth}
\centering
\hspace*{-0mm}\includegraphics[width=1.0\linewidth]{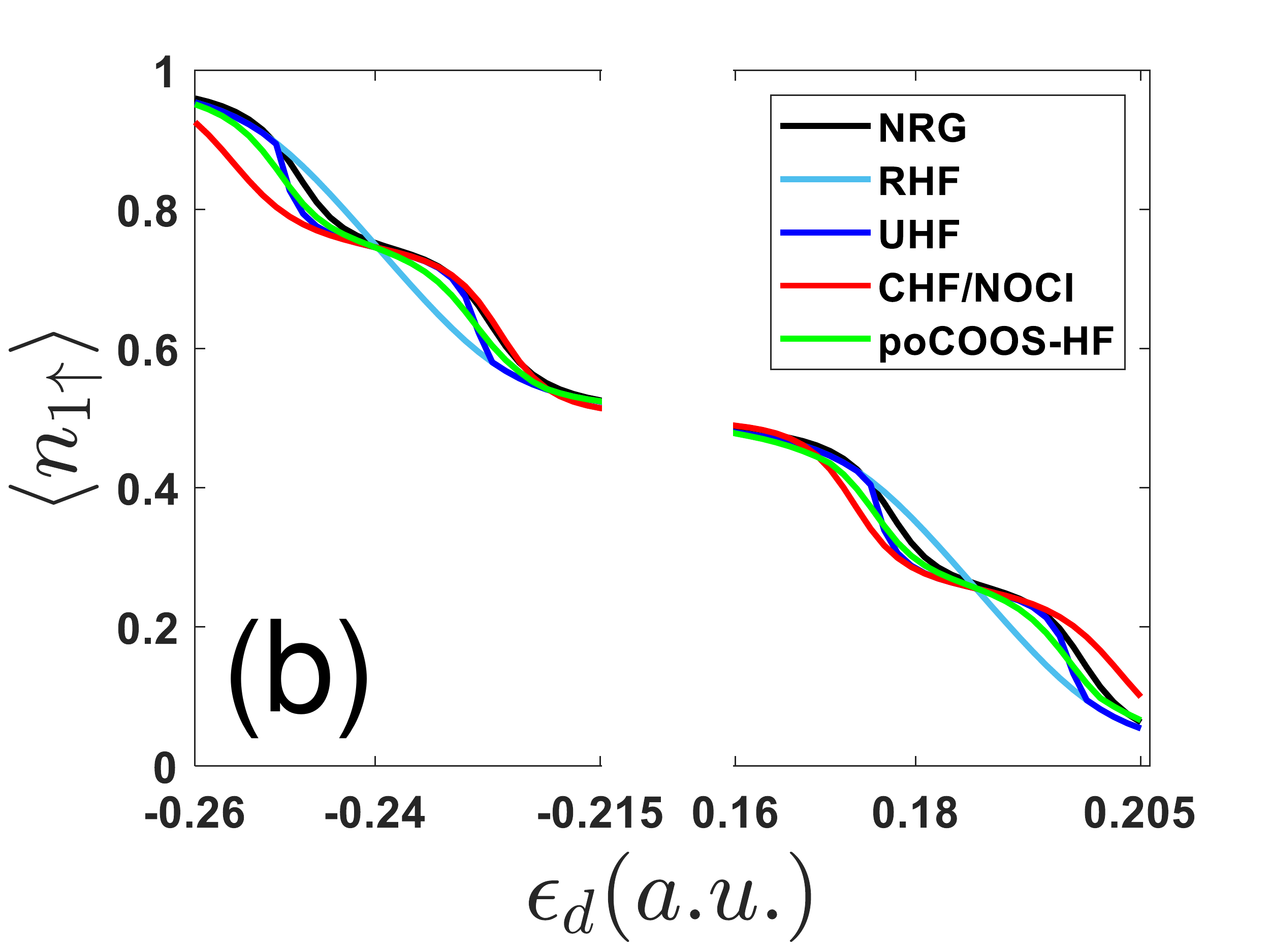}
\end{subfigure}
\end{figure}
\begin{figure}\ContinuedFloat
\begin{subfigure}[t]{\textwidth}
\centering
\vspace{-70pt}
\hspace*{-0mm}\includegraphics[width=1.0\linewidth]{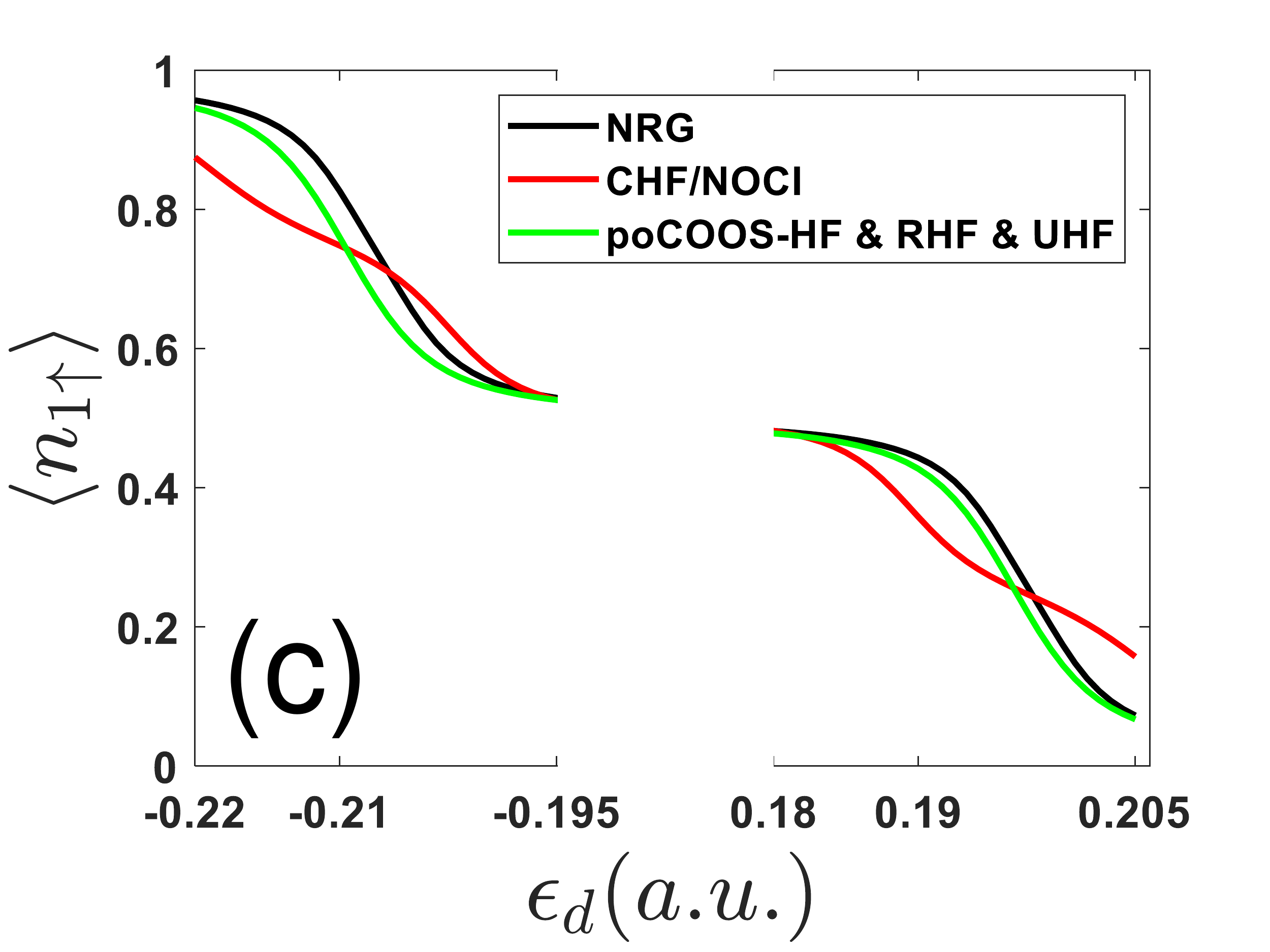}
\end{subfigure}
\begin{minipage}[t]{\textwidth}
\caption{The ground state spin-up electron population on impurity site 1 as a function of the onsite energy $\epsilon_d = \epsilon_{d_1} =\epsilon_{d_2}$, with different onsite repulsion energies. (a) $U=10\Gamma$; (b) $U=5\Gamma$; (c) $U=\Gamma$. Note that poCOOS-HF results match NRG results well for all three different onsite repulsion $U$ regimes. Here, the parameters are: $t_d=0.2$, 803 number of states (801 bath states plus 2 impurities) with band width 0.8 and $\Gamma=0.01$.} 
\label{fig:n_imp}
\end{minipage}

\end{figure}

\begin{figure}[htbp]
\centering
\vspace{-50pt}
\begin{subfigure}[t]{\textwidth}
\centering

\hspace*{-0mm}\includegraphics[width=1.0\linewidth]{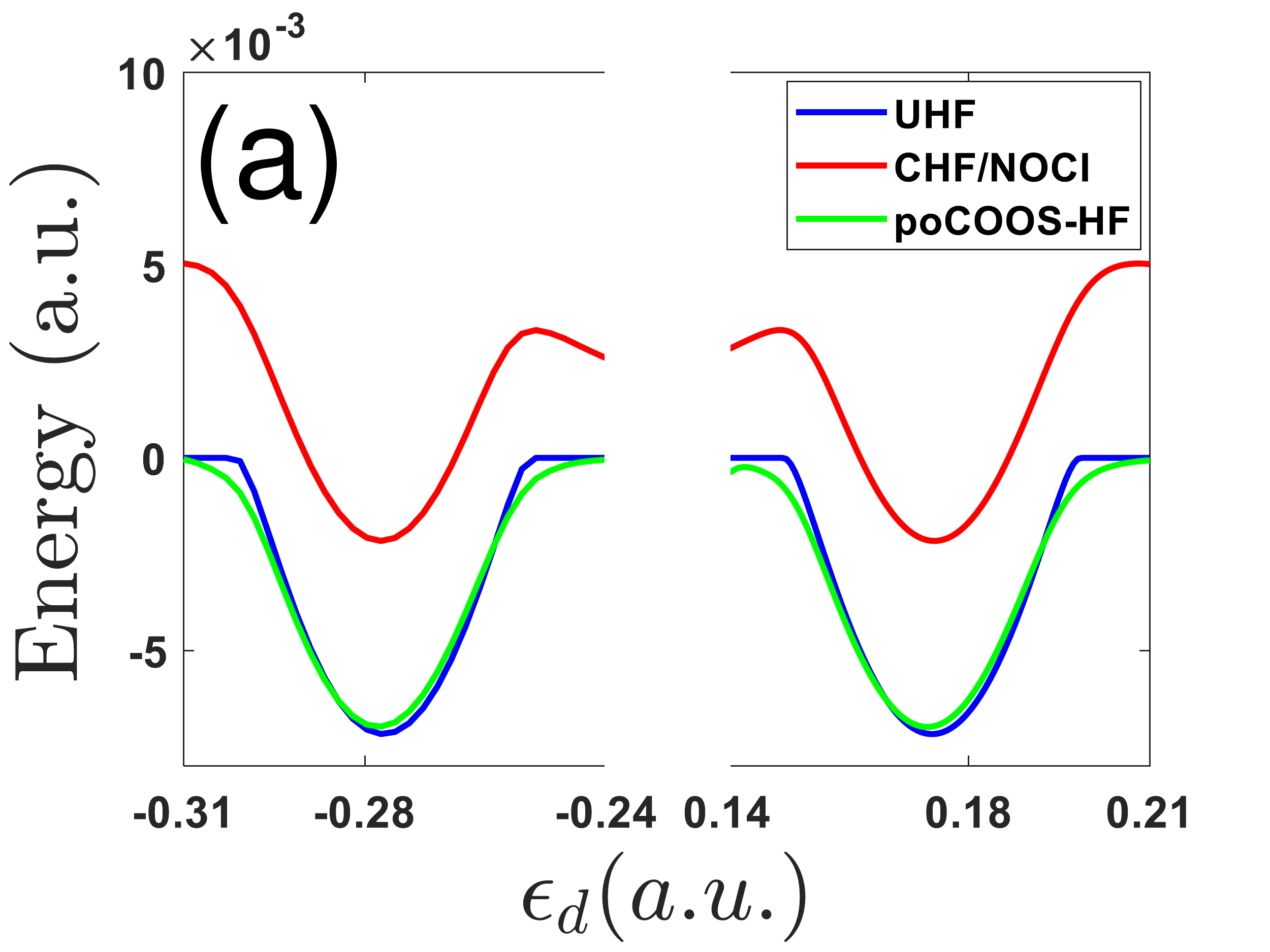}
\end{subfigure}
\begin{subfigure}[t]{\textwidth}
\centering
\hspace*{-0mm}\includegraphics[width=1.0\linewidth]{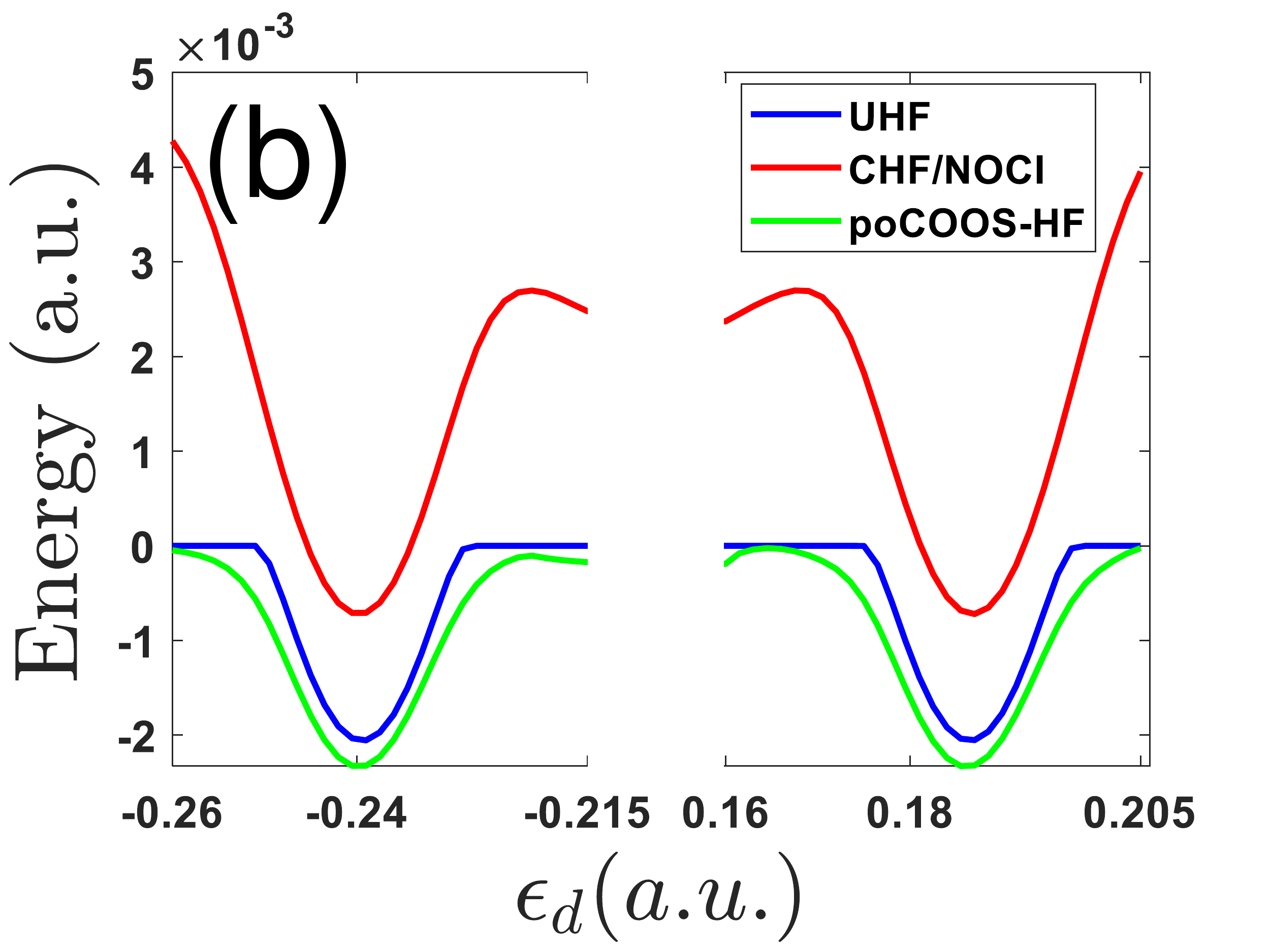}
\end{subfigure}
\end{figure}
\begin{figure}\ContinuedFloat
\begin{subfigure}[t]{\textwidth}
\centering
\vspace{-70pt}
\hspace*{-0mm}\includegraphics[width=1.0\linewidth]{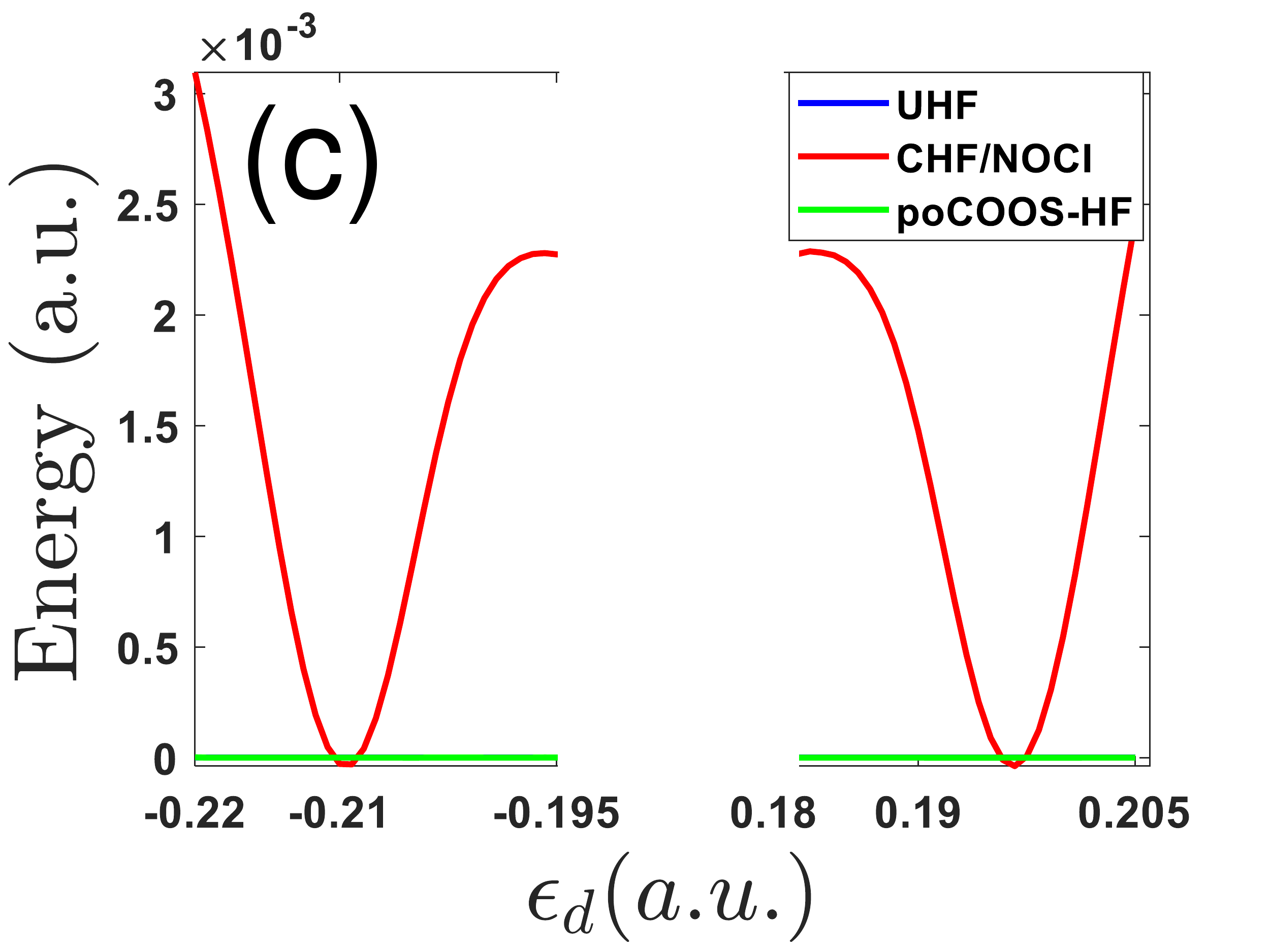}
\end{subfigure}
\begin{minipage}[t]{\textwidth}
\caption{The ground state energy (relative to RHF energy) as a function of the onsite energy $\epsilon_d$, with different onsite repulsion energies. (a) $U=10\Gamma$; (b) $U=5\Gamma$; (c) $U=\Gamma$. Note that the poCOOS-HF energy is comparable to the UHF energy for all three different onsite repulsion $U$ regimes. All parameters are the same as in Fig. \ref{fig:n_imp}.}
\label{fig:energy}
\end{minipage}
\end{figure}


\section{Discussion}
\label{sec:discussion}
\subsection{Spin-Contamination}
\label{subsec:spincontamination}
\begin{figure}[htbp]
\centering
\vspace{-0pt}
\begin{subfigure}[t]{.4\textwidth}
\centering

\hspace*{-40mm}\includegraphics[width=1.5\linewidth]{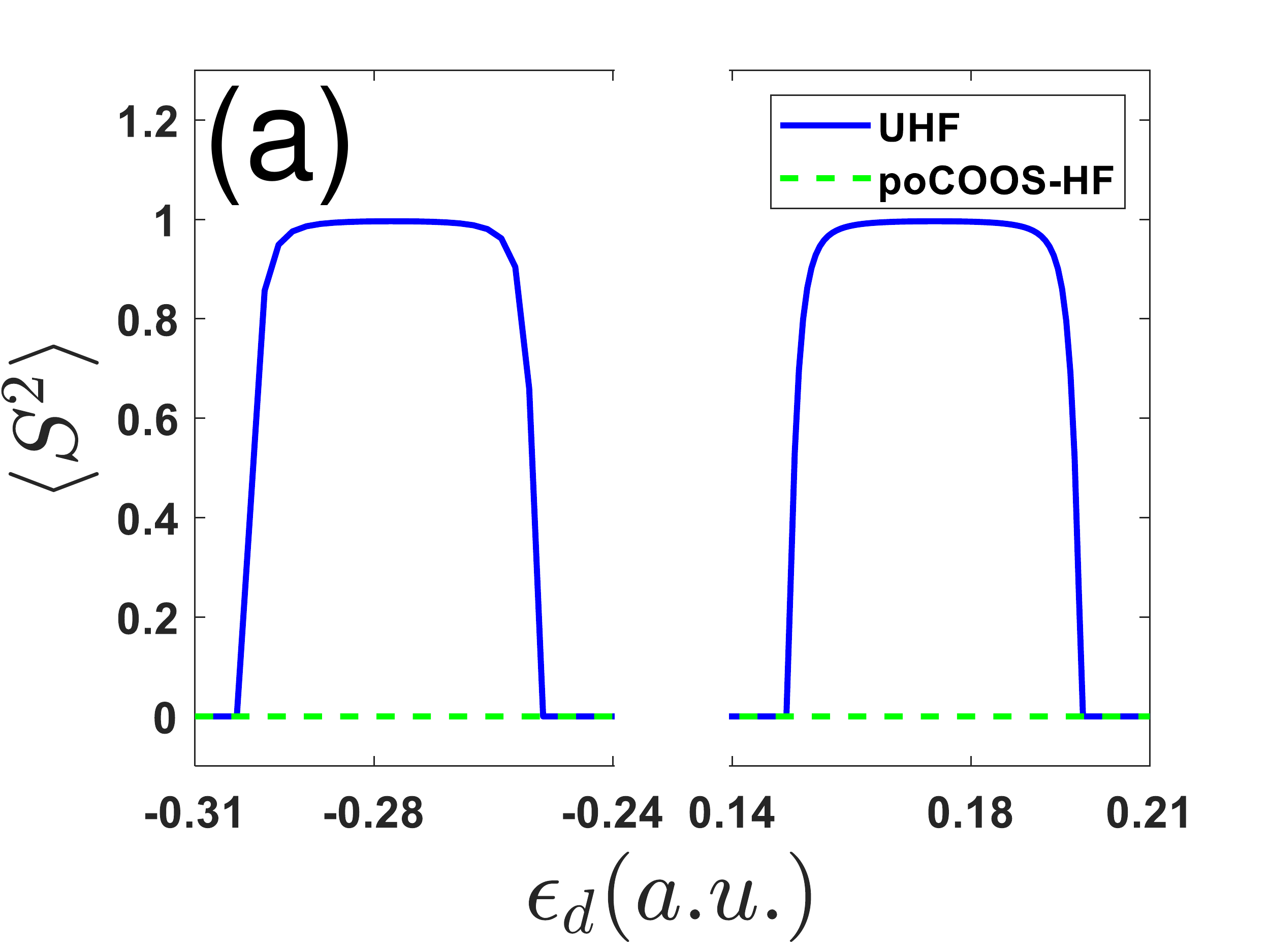}
\end{subfigure}
\begin{subfigure}[t]{.4\textwidth}
\centering
\hspace*{-0mm}\includegraphics[width=1.5\linewidth]{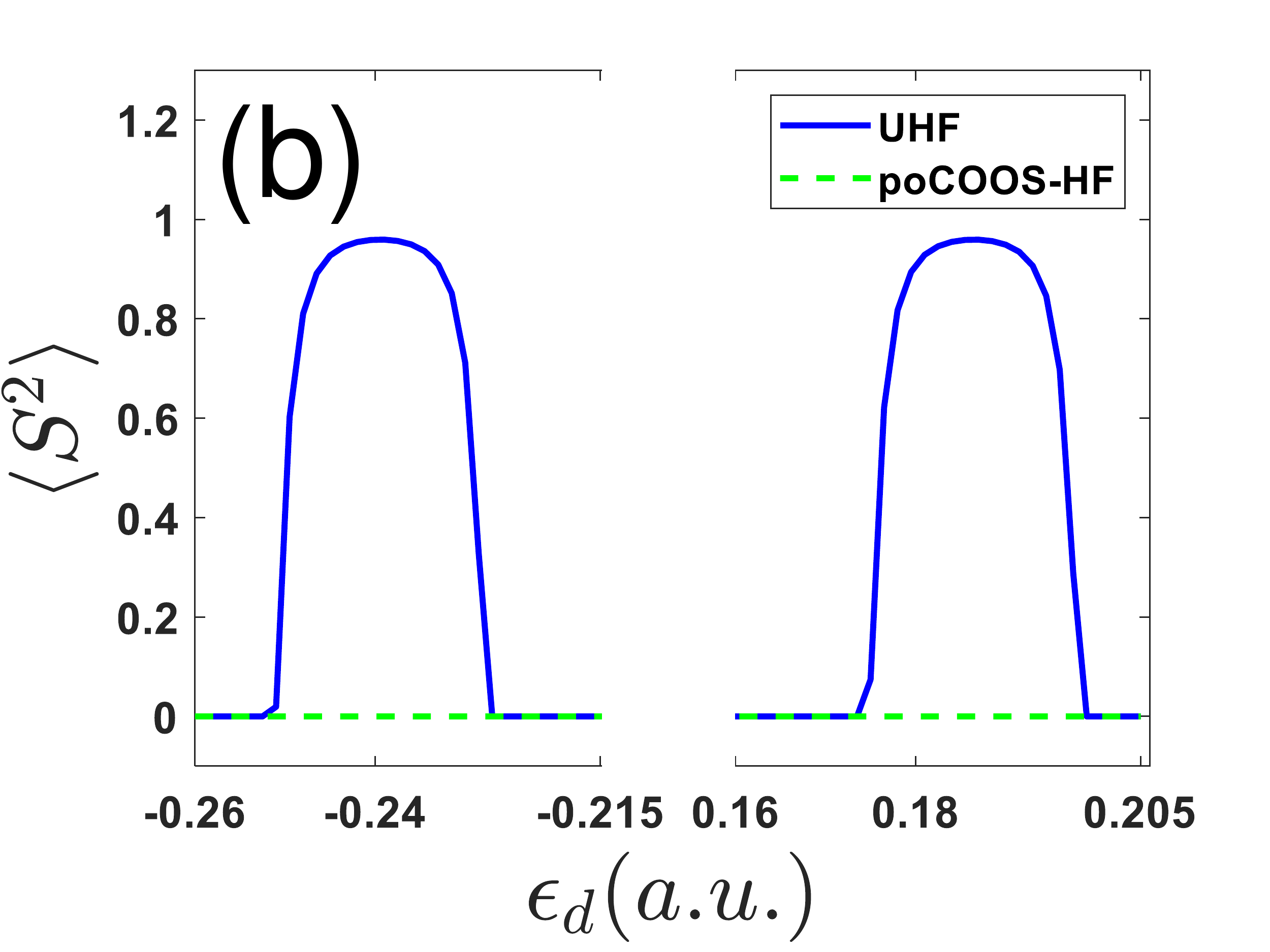}
\end{subfigure}

\medskip

\begin{subfigure}[t]{.4\textwidth}
\centering
\vspace{-0pt}
\hspace*{-40mm}\includegraphics[width=1.5\linewidth]{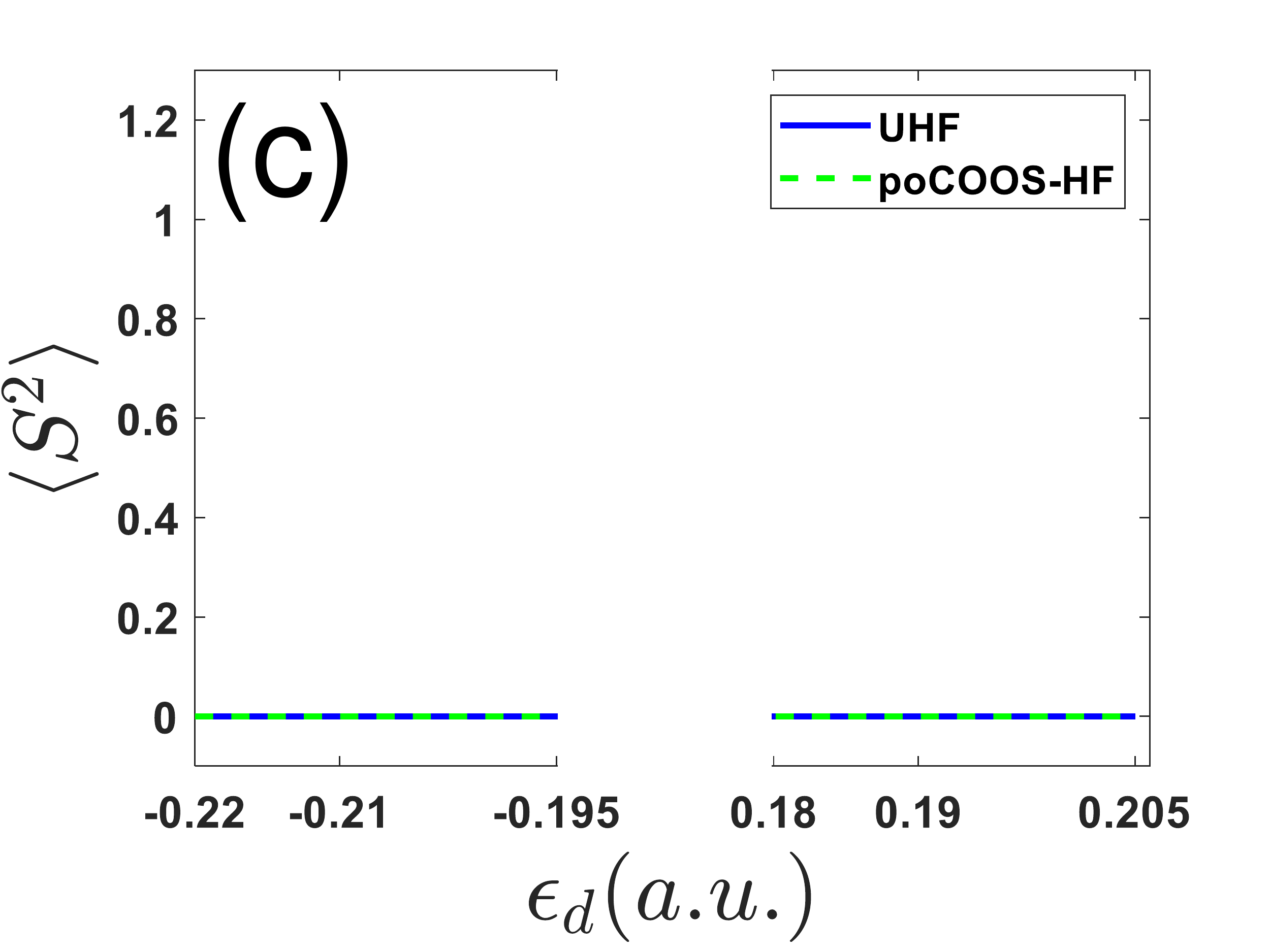}
\end{subfigure}
\begin{minipage}[t]{.4\textwidth}
\caption{Spin contamination ($\expval{S^2}$) as a function of the onsite energy $\epsilon_d$, with different onsite repulsion energies. (a) $U=10\Gamma$; (b) $U=5\Gamma$; (c) $U=\Gamma$. Note that UHF has substantial spin-contamination for $U=5\Gamma$ and $U=10\Gamma$. Also note that poCOOS-HF has no spin-contamination because the algorithm works with a pure singlet wavefunction, see Eq. \ref{eqn:pocooshf}. All parameters are the same as  in Fig. \ref{fig:n_imp}}
\label{fig:S2}
\end{minipage}
\end{figure}

One of the strengths of the poCOOS-HF method is that the method works with a singlet wavefunction, and therefore the method  does not have any spin-contamination.   By contrast, as is well known, UHF does suffer from substantial spin contamination. 
In Fig. \ref{fig:S2}, we report spin-contamination ($\expval{S^2}$) for the three different onsite repulsions $U$. One can see that for $U=5\Gamma$ and $U=10\Gamma$,  $\expval{S^2}$ for UHF can be as large as one; large changes in spin contamination arise near the UHF Coulson-Fisher points.

As a practical matter, the poCOOS-HF ansatz was designed to avoid the two problems just listed.  Our goal was to seek the simplest  spin-pure wavefunction method that could break symmetry (and introduce multi-reference character) in the weak and intermediate coupling, but at the same time recover symmetry in strong coupling limit, all the while being as smooth as possible. For that goal, the two standard candidates in the literature would be: broken symmetry UHF (BS-UHF) and spin-flip (SF) methods. 
It will be interesting to benchmark the results here versus those approaches in the future.
However, already we know that the BS-UHF method is not free of spin contamination.  As for spin-flip methods, often one must include  more configurations than a standard  SF-CIS calculation\cite{alguire2011diabatic, krylov2001spin, shao2003spin} if one wishes to recover a solution free of spin-contamination, e.g. one must implement the SA-SF-CIS method of Zhang and Herbert \cite{zhang2015spin} or the SF-XCIS method of Casanova and Head-Gordon.\cite{casanova2008spin}.  Another interesting option to explore in the future would be Holomorphic Hartree-Fock \cite{hiscock2014holomorphic,burton2016holomorphic} theory where smooth energy curves can obtained, at least for systems not too large.


\subsection{poCOOS vs foCOOS vs CASSCF(2,2)}
At this point, we have seen the poCOOS approach can perform fairly well for the Anderson embedding problem. That being said, one can argue that poCOOS-HF is only a partially optimized theory. After all, the algorithm optimizes only 3-orbitals ($\bm{o,p,q}$ in \ref{eqn:opq}) while the remaining orbitals are kept in the RHF reference state.  In particular,
poCOOS-HF optimizes the orbital $\bm{o}$ within the RHF occupied space, and the remaining core orbitals are not fully optimized in terms of energy as a function orbital rotation parameters (i.e. $\kappa_{it}$ and $\kappa_{ia}$, where $i$ indexes core orbitals, $t$ indexes active orbitals and $a$ indexes virtual orbitals). 

\begin{figure}[htbp]
\centering

\begin{subfigure}[t]{0.4\textwidth}
\centering
\hspace*{-30mm}\includegraphics[width=1.5\linewidth]{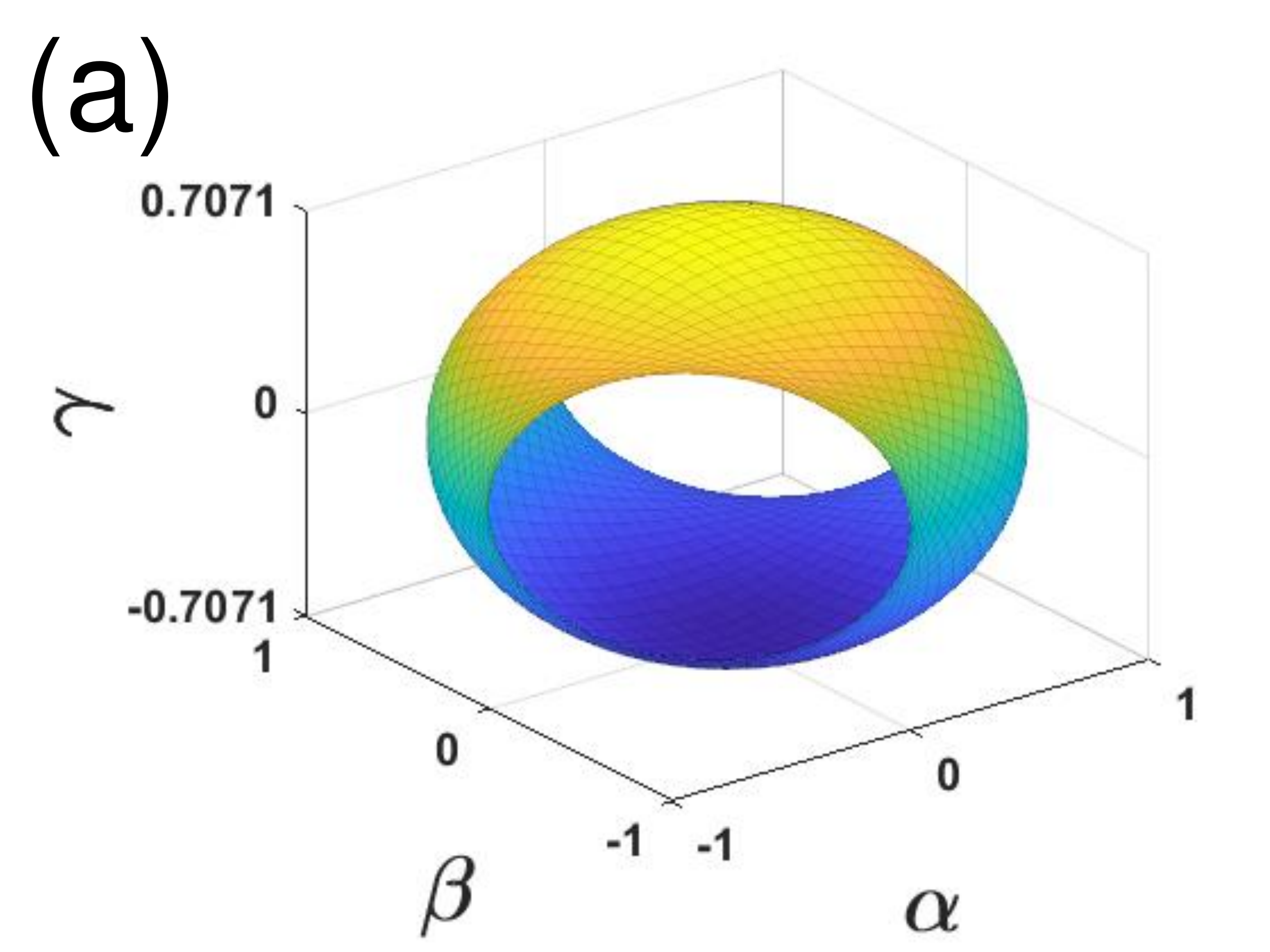}
\end{subfigure}
\begin{subfigure}[t]{0.4\textwidth}
\centering
\hspace*{-0mm}\includegraphics[width=1.5\linewidth]{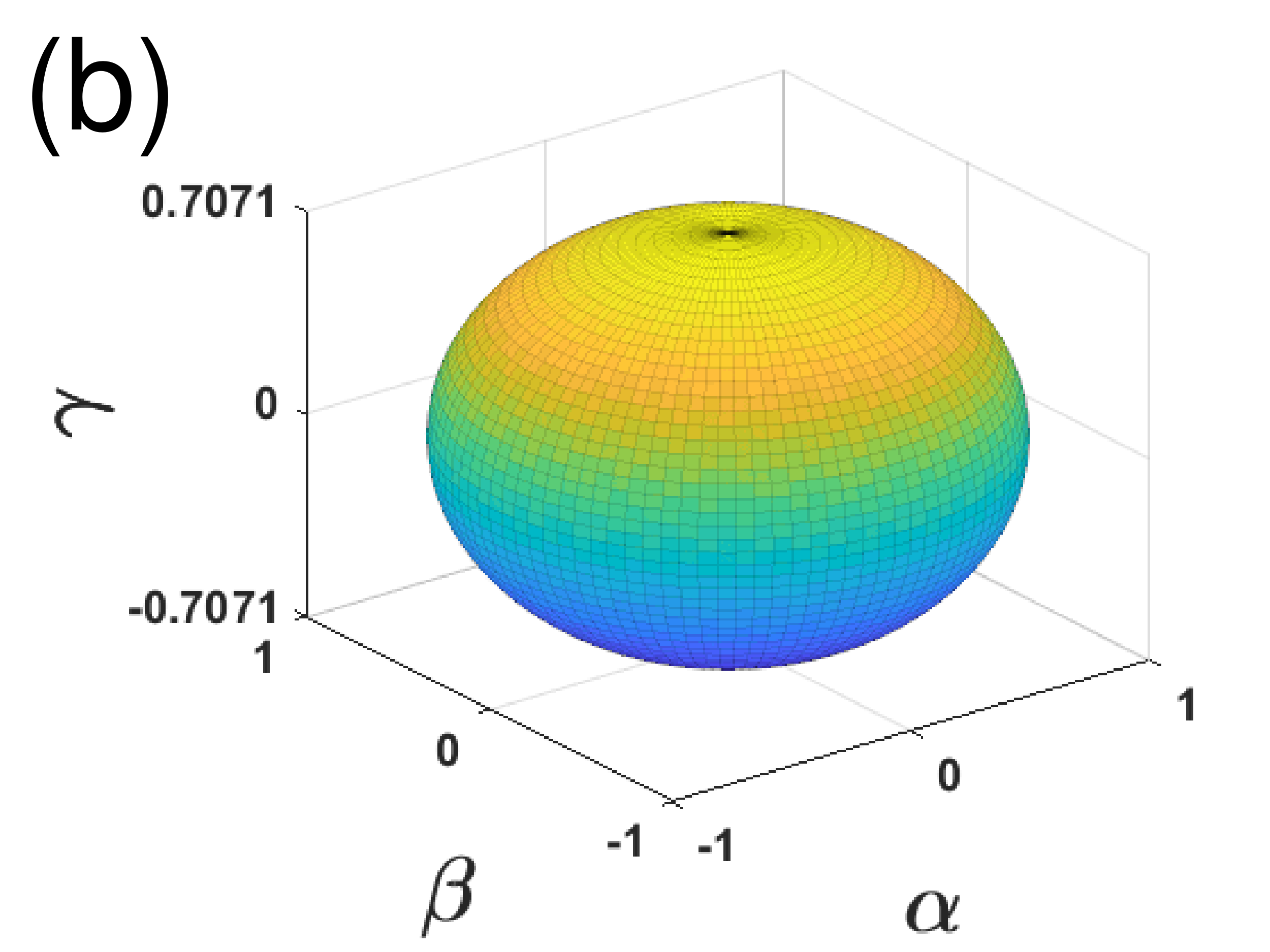}
\end{subfigure}
\caption{CI coefficients parameter space for (a) COOS-HF; (b) CASSCF(2,2). Note that the CASSCF(2,2) parameter space forms a complete ellipsoid ($\alpha^2+\beta^2+2\gamma^2=1$) with semi-axes of (1,1,$\frac{1}{\sqrt{2}}$) while COOS-HF covers just a part of this ellipsoid. The forbidden part of the COOS-HF parameter space corresponds to the areas $\{\alpha>0,\beta>0\}$ and $\{\alpha<0,\beta<0\}$.}
\label{fig:coosvscas}
\end{figure}

In the future, one can imagine fully optimizing all of the orbitals in the  wavefunction of Eq. \ref{eqn:cooshf}, what we might call a \underline{f}ully \underline{o}ptimized foCOOS-HF (see Eq. \ref{eqn:cooshf}
, which optimizes a set of orthogonal orbitals (parametrized by $\kappa_{rs}$, where $r,s$ index any orbitals) and CI coefficients (parametrized by $\alpha, \beta, \gamma$ in Eq. \ref{eqn:CI coeffs}). 
In this spirit, one can clearly ascertain that the foCOOS ansatz would generate many (but not all) configurations suggested by a CASSCF(2,2) wavefunction.  Note that this equivalence is not exact because the parameter space for CASSCF(2,2) CI coefficients is a complete ellipsoid while the parameter space for foCOOS-HF is only a subset of such a  CASSCF(2,2) space. To see this contraint clearly,  consider Eq. \ref{eqn:CI coeffs}, where the normalization constaint can be written as:
\begin{equation}
    \alpha^2+\beta^2+2\gamma^2=1
\end{equation}
This equation can be recast as:
\begin{equation}
    (\alpha+\beta)^2+2\gamma^2-2\alpha\beta=1
\end{equation}
Note that:
\begin{equation}
    2\gamma^2-2\alpha\beta=2\frac{\cos^2\theta\sin^2\eta+\sin^2\theta\cos^2\eta}{N}\ge0,
\end{equation}
so that our COOS-HF ansatz must satisfy:
\begin{equation}
    (\alpha+\beta)^2\leq1
\end{equation}
See Fig. \ref{fig:coosvscas}. Thus, one can consider the parameter space for foCOOS as a ellipsoid cut by two planes, $\alpha+\beta=-1$ and $\alpha+\beta=1$, that is clearly only a subset of the CAS space.  For instance, if one considers the CAS parameter set $\{\alpha=\beta=\frac{1}{\sqrt{2}},\gamma=0\}$, 
clearly there is no  corresponding $\{\theta,\eta\}$ COOS-HF parameter set.

While one might argue that this limitation represents a failure of the COOS-HF wavefunction (because a bigger variational space is always better), we are hopeful that this will not be the case. First, because we have fewer degrees of freedom, we are hopeful that a foCOOS-HF wavefunction (as constructed exclusively from a set of meaningful orbitals), we will be able to build a balanced reference for ground and excited state calculations {\em without state averaging.}   Second, we are also hopeful that using our choice of frontier orbitals, future work with non-orthogonal configuration interaction Hamiltonians (for excited states) will require smaller diagonalizations and few multireference problems.
Third, we are also hopeful that with fewer degrees of freedom, there will be fewer discontinuities to deal with dynamically (and in particular, the code can be run without manually choosing an active space).

As an example of a situation we might expect to encounter,  consider the case where we expect the wavefunction to have half occupation for highest occupied molecular orbital (HOMO) and a half occupation for the lowest unoccupied molecular orbital (LUMO), in other words a one-electron density matrix $D$ of the form: 
\begin{equation}
    D=\sum_{i\in\text{core}}\ket{i}\bra{i}+\frac{1}{2}\ket{h}\bra{h}+\frac{1}{2}\ket{l}\bra{l}
\end{equation}
In order to represent such a density matrix within a CAS parameter space, there are two possible wavefunctions: $\{\alpha=\beta=0,\gamma=\frac{1}{\sqrt{2}}\}$ ( $\frac{1}{\sqrt{2}}(\ket{h\bar{l}}+\ket{l\bar{h}})$) 
 and $\{\alpha=\beta=\frac{1}{\sqrt{2}},\gamma=0\}$ ($\frac{1}{\sqrt{2}}(\ket{h\bar{h}}+\ket{l\bar{l}})$). In the future, it will be interesting to check if CAS calcualtions converge to the same minimum starting from these two different starting guesses. That being said, since the parameter set $\{\alpha=\beta=\frac{1}{\sqrt{2}},\gamma=0\}$ is out of the COOS-HF parameter space, foCOOS will only have one unique guess wavefunction, i.e. $\frac{1}{\sqrt{2}}(\ket{h\bar{l}}+\ket{l\bar{h}})$, and so is less likely to suffer  multiple solutions. In short, the essence of the present approach is that we are willing to deal with less accuracy (i.e. not including a full CAS space) if our goal is really to run nonadiabatic dynamics over a smooth and qualitatively correct surface.
This same reasoning explains why many dynamicists over prefer CASCI to CASSCF for many applications
\cite{levine2021cas}.

\section{Discussion and Conclusion}
\label{sec:conclusion}
In summary, we have presented a straightforward closed or open shell HF approach for electronic structure theory and we have compared such an approach against a CHF/NOCI ansatz for the two-site Anderson model of a molecule sitting on a metal surface. We have tested the algorithms in the  weak ($U=10\Gamma$), intermediate ($U=5\Gamma$) and strong ($U=\Gamma$) coupling regimes and we have found that poCOOS-HF can obtain accurate results as compared with exact numerical renormalization group (NRG) theory, recovering charge transfer states where appropriate; in particular, in the weak and intermediate coupling regimes, poCOOS-HF can recover the strong open-shell singlet character (when $\expval{n_{tot}}=3$ and $\expval{n_{tot}}=1$) exhibited by the impurity. 

Looking forward, the COOS-HF approach has the attractive feature that the algorithm can be completely characterized by a set of orbitals, i.e. one does not need to list any CI coefficients.  As such, the ansatz may prove amenable to a merger with DFT.  Moreover,  in the future, one would like to run molecule non-adiabatic dynamics on metal surfaces through the current framework.  Progress on this front will depend on two important future developments. First, if one looks very carefully at Fig. \ref{fig:n_imp}(a), in the strong $U$ regime, poCOOS-HF would appear to have have four small discontinuities around $\epsilon=-0.30,-0.25,0.15,0.20$, attributable to a Coulson-Fischer point (similar to UHF). The changes in slope are clearly far smaller for poCOOS-HF than for UHF, but they may be unavoidable. One would hope that, if one implements the fo-COOS-HF approach, such a discontinuity will be removed entirely. Second, in the future, it will be essential to generate excited states on top of a COOS-HF reference. Again, it will be essential to generate smooth surfaces as much as possible. Despite these concerns, all indications are that the present approach has the potential to be applied to reasonably sized electronic subsystems interacting with large electronic baths for use in future dynamical approaches.

\begin{acknowledgement}

This work was supported by the U.S. Air Force Office of Scientific Research (USAFOSR) under Grant Nos. FA9550-18-1-0497 and FA9550-18-1-0420. We thank the DoD High Performance Computing Modernization Program for computer time. And we also thank the Texas Advanced Computing Center (TACC) at The University of Texas at Austin for providing HPC resources that have contributed to the research results reported in this paper. URL: http://www.tacc.utexas.edu

\end{acknowledgement}

\section{Appendix} 
\label{sec:appendix}
\subsection{Non-Orthogonal Configuration Matrix Element}
\label{sec:nomel}
Under the framework of constrained Hartree-Fock, suppose we find two ground states with the total number of electrons on the impurities equal to 0 and 1, denoted as $\mel{\Phi_g}{\sum_\sigma d_{1\sigma}^\dagger d_{1\sigma}+d_{2\sigma}^\dagger d_{2\sigma}}{\Phi_g}=1$ and $\mel{\Phi_g'}{\sum_\sigma d_{1\sigma}^\dagger d_{1\sigma}+d_{2\sigma}^\dagger d_{2\sigma}}{\Phi_g'}=0$. The quantity that we want to calculate is the matrix element: $\mel{\Phi_g}{H}{\Phi_g'}$. In the following paragraph, the one electron operator $d^\dagger b$ will be used to derive such non-orthogonal matrix elements; here, $d^\dagger$ represents the impurity atomic orbital and $b^\dagger$ represents the bath atomic orbital. 

The matrix element we wish to calculate can be explicitly written as: $\mel{\Phi_g}{\sum_b\sum_\sigma d_{1\sigma}^\dagger b_\sigma + c.c.}{\Phi_g'}$. For restricted orbitals, we can ignore the summation over spin $\sigma$ and focus on $\mel{\Phi_g}{2\sum_b d_{1}^\dagger b + c.c.}{\Phi_g'}$: 

\begin{equation}
\label{eqn:gg'db}
    \mel{\Phi_g}{\sum_b d_{1}^\dagger b}{\Phi_g'}
    =\mel{\Phi_g}{\sum_b d_{1}^\dagger b}{\Phi_g}\braket{\Phi_g}{\Phi_g'}+\sum_{ia}\mel{\Phi_g}{\sum_b d_{1}^\dagger b}{\Phi_i^a}\braket{\Phi_i^a}{\Phi_g'}
\end{equation}

First of all, using second quantization, it is clear that:

\begin{equation}
\label{eqn:ggdb}
    \mel{\Phi_g}{\sum_b d_{1}^\dagger b}{\Phi_g}=\sum_b\sum_i^{occ}\braket{d_1}{i}\braket{i}{b}
\end{equation}
\begin{equation}
\label{eqn:gsdb}
    \mel{\Phi_g}{\sum_b d_{1}^\dagger b}{\Phi_i^a}=\sum_b\braket{d_1}{i}\braket{a}{b}
\end{equation}

Secondly, let us calculate the overlap:

\begin{equation}
\label{eqn:ggovl}
    \braket{\Phi_g}{\Phi_g'}=\det(S^o)
\end{equation}

Here, $S^o$ is the $N_o$ by $N_o$ occupied orbital overlap matrix (where $N_o$ is the number of occupied orbitals). To calculate the overlap $\braket{\Phi_i^a}{\Phi_g'}$, we need  to introduce a  biorthogonal basis set. If we perform a singular value decomposition on the (occupied-occupied) $S^o$ and (virtual-virtual) $S^v$ overlap matrices,

\begin{equation}
    S^o=U^o\Lambda^oV^{o\dagger}
\end{equation}
\begin{equation}
    S^v=U^v\Lambda^vV^{v\dagger},
\end{equation}

then the molecular orbitals can be expressed in this biorthogonal basis set:

\begin{equation}
    \ket{i}=\sum_k^{occ}\ket{\tilde{k}}U_{ik}^{o*}
\end{equation}
\begin{equation}
    \ket{a}=\sum_c^{vir}\ket{\tilde{c}}U_{ac}^{v*}
\end{equation}
\begin{equation}
    \ket{i'}=\sum_k^{occ}\ket{\tilde{k'}}V_{ik}^{o*}
\end{equation}
\begin{equation}
    \ket{a'}=\sum_c^{vir}\ket{\tilde{c'}}V_{ac}^{v*}
\end{equation}

We can then express the overlap in this basis of biorthogonal orbitals:

\begin{equation}
\label{eqn:sgovl}
    \braket{\Phi_i^a}{\Phi_g'}=\sum_{k}^{occ}\sum_c^{vir}U_{ac}^{v}U_{ik}^{o*}\braket{\Phi_{\tilde{k}}^{\tilde{c}}}{\Phi_{\tilde{g}}'}=\sum_{k}^{occ}\sum_c^{vir}U_{ac}^{v}U_{ik}^{o*}\frac{\det(S^o)\braket{\tilde{c}}{\tilde{k}'}}{\lambda_k}=\sum_{k}^{occ}U_{ik}^{o*}\frac{\det(S^o)\braket{a}{\tilde{k}'}}{\lambda_k}
\end{equation}
The second equality uses the fact that  $det(S^o)=\braket{\Phi_g}{\Phi_g'}=\braket{\Phi_{\tilde{g}}}{\Phi_{\tilde{g}}'}=\Pi_k^{occ}\lambda_k$ with $\lambda_k$ being the diagonal element of $\Lambda^o$. 

Next, the expression above can be simplified again using the identity:
\begin{equation}
\label{eqn:Sinv}
\begin{split}
    \sum_k^{occ}U_{ik}^{o*}\frac{1}{\lambda_k}\ket{\tilde{k}'}
    &=\sum_{kl}^{occ}U_{ik}^{o*}\frac{\delta_{kl}}{\lambda_k}\ket{\tilde{l}'}=\sum_{klm}^{occ}U_{ik}^{o*}\frac{1}{\lambda_k}(V^{o*\dagger})_{km}V^{o*}_{ml}\ket{\tilde{l}'}=\sum_{km}^{occ}U_{ik}^{o*}\frac{1}{\lambda_k}(V^{o*\dagger})_{km}\ket{m'}\\
    &=\sum_{m}^{occ}(V^o\frac{1}{\Lambda^o}U^{o\dagger})_{mi}\ket{m'}=\sum_{m}^{occ}(S^o)^{-1}_{mi}\ket{m'}
    \end{split}
\end{equation}

If we substitute Eq.\ref{eqn:Sinv} into Eq.\ref{eqn:sgovl}, we find:

\begin{equation}
\label{eqn:sgovlf}
    \braket{\Phi_i^a}{\Phi_g'}=\det(S^o)\sum_{m}^{occ}(S^o)^{-1}_{mi}\braket{a}{m'}
\end{equation}

Finally, by combining Eq.\ref{eqn:ggovl} and Eq.\ref{eqn:ggdb}, we evaluate the first term of Eq.\ref{eqn:gg'db}.

\begin{equation}
\label{eqn:gg'db1}
    \mel{\Phi_g}{\sum_b d_{1}^\dagger b}{\Phi_g}\braket{\Phi_g}{\Phi_g'}=\det(S^o)\sum_{b\in bath}\sum_i^{occ}\braket{d_1}{i}\braket{i}{b}
\end{equation}

And by combining Eq.\ref{eqn:sgovlf} and Eq.\ref{eqn:gsdb}, we can also evaluate the second term of Eq.\ref{eqn:gg'db}.

\begin{equation}
\label{eqn:gg'db2}
\begin{split}
    \sum_{ia}\mel{\Phi_g}{\sum_b d_{1}^\dagger b}{\Phi_i^a}\braket{\Phi_i^a}{\Phi_g'}
    &=\det(S^o)\sum_{ia}\sum_b\braket{d_1}{i}\braket{a}{b}\sum_{m}^{occ}(S^o)^{-1}_{mi}\braket{a}{m'}\\
    &=\det(S^o)\sum_b\sum_{ia}\sum_{m}^{occ}\braket{b}{a}\braket{a}{m'}(S^o)^{-1}_{mi}\braket{i}{d_1}\\
    &=\det(S^o)\sum_b\sum_{i}\sum_{m}^{occ}\braket{b}{m'}(S^o)^{-1}_{mi}\braket{i}{d_1}\\
    &-\det(S^o)\sum_b\sum_{i}^{occ}\braket{b}{i}\braket{i}{d_1}
    \end{split}
\end{equation}

The third equality uses the property that

\begin{equation}
    \sum_a^{vir}\ketbra{a}{a}=I-\sum_j^{occ}\ketbra{j}{j}
\end{equation}
\begin{equation}
    \sum_m^{occ}\braket{j}{m'}(S^o)^{-1}_{mi}=\sum_m^{occ}(S^o)_{jm}(S^o)^{-1}_{mi}=\delta_{ji}
\end{equation}

Altogether, by substituting Eq.\ref{eqn:gg'db1} and Eq.\ref{eqn:gg'db2} into Eq.\ref{eqn:gg'db}, we recover

\begin{equation}
    \mel{\Phi_g}{\sum_b d_{1}^\dagger b}{\Phi_g'}=\det(S^o)\sum_{b\in bath}\sum_{im}^{occ}\braket{b}{m'}(S^o)^{-1}_{mi}\braket{i}{d_1}
\end{equation}
Now we turn to the two-electron operator contribution:
\begin{equation}
\begin{aligned}
    \mel{\Phi_g}{d^{\dagger}d\bar{d^{\dagger}}\bar{d}}{\Phi'_g}&=\bra{\Phi_g}d^{\dagger}d\left(\ket{\Phi_g}\bra{\Phi_g}+\ket{\Phi_i^a}\bra{\Phi_i^a}\right)\left(\ket{\Phi'_g}\bra{\Phi'_g}+\ket{\Phi_{j'}^{b'}}\bra{\Phi_{j'}^{b'}}\right)\bar{d^{\dagger}}\bar{d}\ket{\Phi'_g}\\
    &=\mel{\Phi_g}{d^{\dagger}d}{\Phi_g}\braket{\Phi_g}{\Phi'_g}\mel{\Phi'_g}{\bar{d^{\dagger}}\bar{d}}{\Phi'_g}\\
    &+\mel{\Phi_g}{d^{\dagger}d}{\Phi_g}\braket{\Phi_g}{\Phi_{j'}^{b'}}\mel{\Phi_{j'}^{b'}}{\bar{d^{\dagger}}\bar{d}}{\Phi'_g}\\
    &+\mel{\Phi_g}{d^{\dagger}d}{\Phi_i^a}\braket{\Phi_i^a}{\Phi'_g}\mel{\Phi'_g}{\bar{d^{\dagger}}\bar{d}}{\Phi'_g}\\
    &+\mel{\Phi_g}{d^{\dagger}d}{\Phi_i^a}\braket{\Phi_i^a}{\Phi_{j'}^{b'}}\mel{\Phi_{j'}^{b'}}{\bar{d^{\dagger}}\bar{d}}{\Phi'_g}\\
\end{aligned}
\end{equation}
where $i,j,a,b$ are dummy summation indices. Then, we define:
\begin{equation}
    \begin{aligned}
    n_g &= \mel{\Phi_g}{d^{\dagger}d}{\Phi_g}\\
    n_{g'} &= \mel{\Phi_{g'}}{d^{\dagger}d}{\Phi_{g'}}
    \end{aligned}
\end{equation} 
Then, it follows that:
\begin{equation}
    \begin{aligned}
        \mel{\Phi_g}{d^{\dagger}d}{\Phi_g}\braket{\Phi_g}{\Phi'_g}\mel{\Phi'_g}{\bar{d^{\dagger}}\bar{d}}{\Phi'_g}&=\det(S^o)n_gn_{g'}\\
        \sum_{j'}^{occ}\sum_{b'}^{vir}\mel{\Phi_g}{d^{\dagger}d}{\Phi_g}\braket{\Phi_g}{\Phi_{j'}^{b'}}\mel{\Phi_{j'}^{b'}}{\bar{d^{\dagger}}\bar{d}}{\Phi'_g}&=\det(S^o)n_g\sum_{ij'}^{occ}\braket{d}{j'}(S^o)^{-1}_{j'i}\braket{i}{d}-\det(S^o)n_gn_{g'}\\
        \sum_{i}^{occ}\sum_{a}^{vir}\mel{\Phi_g}{d^{\dagger}d}{\Phi_i^a}\braket{\Phi_i^a}{\Phi'_g}\mel{\Phi'_g}{\bar{d^{\dagger}}\bar{d}}{\Phi'_g}&=\det(S^o)n_{g'}\sum_{ij'}^{occ}\braket{d}{j'}(S^o)^{-1}_{j'i}\braket{i}{d}-\det(S^o)n_gn_{g'}\\
        \sum_{ij'}^{occ}\sum_{ab'}^{vir}\mel{\Phi_g}{d^{\dagger}d}{\Phi_i^a}\braket{\Phi_i^a}{\Phi_{j'}^{b'}}\mel{\Phi_{j'}^{b'}}{\bar{d^{\dagger}}\bar{d}}{\Phi'_g}&=\det(S^o)\sum_{im'}^{occ}\braket{d}{m'}(S^o)^{-1}_{m'i}\braket{i}{d}\sum_{kj'}^{occ}\braket{d}{j'}(S^o)^{-1}_{j'k}\braket{k}{d}\\
        &-\det(S^o)n_g\sum_{kj'}^{occ}\braket{d}{j'}(S^o)^{-1}_{j'k}\braket{k}{d}\\
        &-\det(S^o)n_{g'}\sum_{im'}^{occ}\braket{d}{m'}(S^o)^{-1}_{m'i}\braket{i}{d}\\
        &+\det(S^o)n_gn_{g'}
    \end{aligned}
\end{equation}
Altogether, there is a lot of cancellation, and the two-electron operator contribution is:
\begin{equation}
    \mel{\Phi_g}{d^{\dagger}d\bar{d^{\dagger}}\bar{d}}{\Phi'_g}=\det(S^o)\sum_{im'}^{occ}\braket{d}{m'}(S^o)^{-1}_{m'i}\braket{i}{d}\sum_{kj'}^{occ}\braket{d}{j'}(S^o)^{-1}_{j'k}\braket{k}{d}
\end{equation}
\subsection{Analytical Energy Gradient for poCOOS-HF}
\label{sec:gradient}
Lastly, for  the sake of completeness, here we list all of the electronic derivatives (with respect to orbital variations) of the relevant fock operators, overlap matrix elements, and two electron matrix elements as present in Eq. \ref{eqn:Epocoos}.
\subsubsection{Derivatives of Fock Operators}
\begin{equation}
\pdv{f_{oo}}{c_i}=2\sum_jc_jf_{ij}
\end{equation}
\begin{equation}
\begin{aligned}
\pdv{f_{pp}}{c_i}&=d_o^2\cdot2\sum_jc_jf_{ij}+2\sum_ad_od_af_{ia}\\
\pdv{f_{pp}}{d_o}&=2d_o\sum_{ij}c_ic_jf_{ij}+2\sum_ad_a\sum_ic_if_{ia}\\
\pdv{f_{pp}}{d_a}&=2d_o\sum_ic_if_{ia}+2\sum_bd_bf_{ab}
\end{aligned}
\end{equation}
\begin{equation}
\begin{aligned}
\pdv{f_{qq}}{c_i}&=\tilde{d}_o^2\cdot2\sum_jc_jf_{ij}+2\sum_a\tilde{d}_o\tilde{d}_af_{ia}\\
\pdv{f_{qq}}{\tilde{d}_o}&=2\tilde{d}_o\sum_{ij}c_ic_jf_{ij}+2\sum_a\tilde{d}_a\sum_ic_if_{ia}\\
\pdv{f_{qq}}{\tilde{d}_a}&=2\tilde{d}_o\sum_ic_if_{ia}+2\sum_b\tilde{d}_bf_{ab}
\end{aligned}
\end{equation}
\begin{equation}
\begin{aligned}
\pdv{f_{pq}}{c_i}&=d_o\tilde{d}_o\cdot2\sum_jc_jf_{ij}+\sum_a\tilde{d}_od_af_{ia}+\sum_a d_o\tilde{d}_af_{ia}\\
\pdv{f_{pq}}{d_o}&=\tilde{d}_o\sum_{ij}c_ic_jf_{ij}+\sum_a \tilde{d}_a\sum_ic_if_{ia}\\
\pdv{f_{pq}}{\tilde{d}_o}&=d_o\sum_{ij}c_ic_jf_{ij}+\sum_a d_a\sum_ic_if_{ia}\\
\pdv{f_{pq}}{d_a}&=\tilde{d}_o\sum_ic_if_{ia}+\sum_b\tilde{d}_bf_{ab}\\
\pdv{f_{pq}}{\tilde{d}_a}&=d_o\sum_ic_if_{ia}+\sum_b d_bf_{ab}
\end{aligned}
\end{equation}
\subsubsection{Derivatives of Overlap Matrices}
\begin{equation}
\begin{aligned}
\pdv{\braket{p}{q}}{c_i}&=d_o\tilde{d}_o\cdot2c_i+\sum_ad_o\tilde{d}_a\braket{i}{a}+\sum_a\tilde{d}_od_a\braket{i}{a}\\
\pdv{\braket{p}{q}}{d_o}&=\tilde{d}_o\sum_ic_i^2+\sum_{ai}\tilde{d}_ac_i\braket{i}{a}\\
\pdv{\braket{p}{q}}{\tilde{d}_o}&=d_o\sum_ic_i^2+\sum_{ai}d_ac_i\braket{i}{a}\\
\pdv{\braket{p}{q}}{d_a}&=\sum_{b}\tilde{d}_b\braket{a}{b}+\sum_i\tilde{d}_oc_i\braket{i}{a}\\
\pdv{\braket{p}{q}}{\tilde{d}_a}&=\sum_{b}d_b\braket{b}{a}+\sum_id_oc_i\braket{i}{a}\\
\end{aligned}
\end{equation}
\subsubsection{Derivatives of Two Electron Integrals}
\begin{equation}
\pdv{(oo|oo)}{c_i}=4\sum_{jkl}c_jc_kc_l(ij|kl)
\end{equation}
\begin{equation}
\begin{aligned}
\pdv{(pq|qp)}{c_i} &=d_o^2\tilde{d}_o^2\sum_{jkl}4c_jc_kc_l(ij|kl)\\
    &+(d_o^2\cdot2\sum_a\tilde{d}_o\tilde{d}_a+2\sum_ad_od_a\cdot\tilde{d}_o^2)\cdot\sum_{jk}3c_jc_k(ij|ka)\\
    &+(d_o^2\cdot\sum_{ab}\tilde{d}_a\tilde{d}_b+2\sum_ad_od_a\cdot2\sum_b\tilde{d}_o\tilde{d}_b+\sum_{ab}d_ad_b\cdot\tilde{d}_o^2)\cdot\sum_{j}2c_j(ij|ab)\\
    &+(2\sum_ad_od_a\cdot\sum_{bc}\tilde{d}_b\tilde{d}_c+\sum_{ab}d_ad_b\cdot2\sum_c\tilde{d}_o\tilde{d}_c)\cdot(ia|bc)
\end{aligned}
\end{equation}
\begin{equation}
\begin{aligned}
 \pdv{(pq|qp)}{d_o}&=2d_o\tilde{d}_o^2\sum_{ijkl}c_ic_jc_kc_l(ij|kl)\\
    &+(2d_o\cdot2\sum_a\tilde{d}_o\tilde{d}_a+2\sum_ad_a\cdot\tilde{d}_o^2)\cdot\sum_{ijk}c_ic_jc_k(ij|ka)\\
    &+(2d_o\cdot\sum_{ab}\tilde{d}_a\tilde{d}_b+2\sum_ad_a\cdot2\sum_b\tilde{d}_o\tilde{d}_b)\cdot\sum_{ij}c_ic_j(ij|ab)\\
    &+(2\sum_ad_a\cdot\sum_{bc}\tilde{d}_b\tilde{d}_c)\cdot\sum_ic_i(ia|bc)
\end{aligned}
\end{equation}
\begin{equation}
\begin{aligned}
    \pdv{(pq|qp)}{d_a}&=(2d_o\cdot\tilde{d}_o^2)\cdot\sum_{ijk}c_ic_jc_k(ij|ka)\\
    &+(2d_o\cdot2\sum_b\tilde{d}_o\tilde{d}_b+\sum_{b}2d_b\cdot\tilde{d}_o^2)\cdot\sum_{ij}c_ic_j(ij|ab)\\
    &+(2d_o\cdot\sum_{bc}\tilde{d}_b\tilde{d}_c+\sum_{b}2d_b\cdot2\sum_c\tilde{d}_o\tilde{d}_c)\cdot\sum_ic_i(ia|bc)\\
&+\sum_{b}2d_b\cdot\sum_{cd}\tilde{d}_c\tilde{d}_d(ab|cd)
\end{aligned}
\end{equation}
\begin{equation}
\begin{aligned}
\pdv{(pq|qp)}{\tilde{d}_o}&=2d_o^2\tilde{d}_o\sum_{ijkl}c_ic_jc_kc_l(ij|kl)\\
    &+(d_o^2\cdot2\sum_a\tilde{d}_a+2\sum_ad_od_a\cdot2\tilde{d}_o)\cdot\sum_{ijk}c_ic_jc_k(ij|ka)\\
    &+(2\sum_ad_od_a\cdot2\sum_b\tilde{d}_b+\sum_{ab}d_ad_b\cdot2\tilde{d}_o)\cdot\sum_{ij}c_ic_j(ij|ab)\\
    &+(\sum_{ab}d_ad_b\cdot2\sum_c\tilde{d}_c)\cdot\sum_ic_i(ia|bc)
\end{aligned}
\end{equation}   
\begin{equation}
\begin{aligned}
\pdv{(pq|qp)}{\tilde{d}_b}&=(d_o^2\cdot2\tilde{d}_o)\cdot\sum_{ijk}c_ic_jc_k(ij|kb)\\
    &+(d_o^2\cdot\sum_{a}2\tilde{d}_a+2\sum_ad_od_a\cdot2\tilde{d}_o)\cdot\sum_{ij}c_ic_j(ij|ab)\\
    &+(2\sum_ad_od_a\cdot\sum_{c}2\tilde{d}_c+\sum_{ac}d_ad_c\cdot2\tilde{d}_o)\cdot\sum_ic_i(ia|cb)\\
&+\sum_{ac}d_ad_c\cdot\sum_{d}2\tilde{d}_d(ac|bd)
\end{aligned}
\end{equation}
\begin{equation}
\begin{aligned}
\pdv{(oo|pp)}{c_i}
&=d_o^2\cdot\sum_{jkl}4c_jc_kc_l(ij|kl)+2\sum_ad_od_a\cdot\sum_{jk}3c_jc_k(ij|ka)+\sum_{ab}d_ad_b\cdot\sum_{j}2c_j(ij|ab)\\
\pdv{(oo|pp)}{d_o}&=2d_o\cdot\sum_{ijkl}c_ic_jc_kc_l(ij|kl)+2\sum_ad_a\cdot\sum_{ijk}c_ic_jc_k(ij|ka)\\
\pdv{(oo|pp)}{d_a}&=2d_o\cdot\sum_{ijk}c_ic_jc_k(ij|ka)+\sum_{b}2d_b\cdot\sum_{ij}c_ic_j(ij|ab)
\end{aligned}
\end{equation}
\begin{equation}
\begin{aligned}
\pdv{(oo|qq)}{c_i}&=\tilde{d}_o^2\cdot\sum_{jkl}4c_jc_kc_l(ij|kl)+2\sum_a\tilde{d}_o\tilde{d}_a\cdot\sum_{jk}3c_jc_k(ij|ka)+\sum_{ab}\tilde{d}_a\tilde{d}_b\cdot\sum_{j}2c_j(ij|ab)\\
\pdv{(oo|qq)}{\tilde{d}_o}&=2\tilde{d}_o\cdot\sum_{ijkl}c_ic_jc_kc_l(ij|kl)+2\sum_a\tilde{d}_a\cdot\sum_{ijk}c_ic_jc_k(ij|ka)\\
\pdv{(oo|qq)}{\tilde{d}_a}&=2\tilde{d}_o\cdot\sum_{ijk}c_ic_jc_k(ij|ka)+\sum_{b}2\tilde{d}_b\cdot\sum_{ij}c_ic_j(ij|ab)
\end{aligned}
\end{equation}
\begin{equation}
\begin{aligned}
\pdv{(oo|pq)}{c_i}&=d_o\tilde{d}_o\cdot\sum_{jkl}4c_ic_jc_kc_l(ij|kl)+\sum_a(d_a\tilde{d}_o+d_o\tilde{d}_a)\cdot\sum_{jk}3c_ic_jc_k(ij|ka)+\sum_{ab}d_a\tilde{d}_b\cdot\sum_{ij}2c_j(ij|ab)\\
\pdv{(oo|pq)}{d_o}&=\tilde{d}_o\cdot\sum_{ijkl}c_ic_jc_kc_l(ij|kl)+\sum_a\tilde{d}_a\cdot\sum_{ijk}c_ic_jc_k(ij|ka)\\
\pdv{(oo|pq)}{\tilde{d}_o}&=d_o\cdot\sum_{ijkl}c_ic_jc_kc_l(ij|kl)+\sum_ad_a\cdot\sum_{ijk}c_ic_jc_k(ij|ka)\\
\pdv{(oo|pq)}{d_a}&=\tilde{d}_o\cdot\sum_{ijk}c_ic_jc_k(ij|ka)+\sum_{b}\tilde{d}_b\cdot\sum_{ij}c_ic_j(ij|ab)\\
\pdv{(oo|pq)}{\tilde{d}_a}&=d_o\cdot\sum_{ijk}c_ic_jc_k(ij|ka)+\sum_{b}d_b\cdot\sum_{ij}c_ic_j(ij|ab)
\end{aligned}
\end{equation}




\bibliography{jctc}

\newpage


\end{document}